\documentclass[10pt,a4paper]{article}

\usepackage{amssymb,amsmath}
\usepackage{graphicx,graphics}
\usepackage{mathtools}
\usepackage{bbold}
\usepackage[english]{babel}
\usepackage[utf8]{inputenc}
\usepackage{epsfig,url}
\usepackage{bbm,theorem}
\usepackage{a4wide}
\usepackage{color}
\usepackage{enumerate}
\usepackage{calrsfs}
\DeclareMathAlphabet{\pazocal}{OMS}{zplm}{m}{n}

\newtheorem{theorem}{Theorem}[section]

\newtheorem{assumption}[theorem]{Assumption}

{\theorembodyfont{\upshape}
\newtheorem{remark}[theorem]{Remark}

}
\numberwithin{equation}{section}
\numberwithin{theorem}{section}

\newcommand{\qed}{\hfill$\Box$}

\newcommand{\E}{{\mathbb E}}
\newcommand{\Z}{{\mathbb Z}}
\newcommand{\R}{{\mathbb R}}
\newcommand{\C}{{\mathbb C\hspace{0.05 ex}}}

\newcommand{\T}{{\mathbb T}}

\newcommand{\cf}{{\mathbbm 1}}

\newcommand{\ci}{{\rm i}}
\newcommand{\re}{{\rm Re\,}}
\newcommand{\im}{{\rm Im\,}}
\newcommand{\rme}{{\rm e}}
\newcommand{\rmd}{{\rm d}}
\newcommand{\diag}{{\rm diag}}

\newcommand{\FT}[1]{\widehat{#1}}

\DeclareMathOperator{\Tr}{Tr}

\newcommand{\norm}[1]{\Vert #1\Vert}
\newcommand{\defset}[2]{ \left\{ #1\left|\,
 #2\makebox[0cm]{$\displaystyle\phantom{#1}$}\right.\!\right\} }

\newcommand{\mean}[1]{\langle #1\rangle}
\newcommand{\vep}{\varepsilon}
\newcommand{\defem}[1]{{\em #1\/}}
\newcommand{\qand}{\quad\text{and}\quad}

\newcounter{jlisti}
\newenvironment{jlist}[1][(\thejlisti)]{\begin{list}{{\rm #1}\ \ }{ %
      \usecounter{jlisti} %
    \setlength{\itemsep}{0pt}
    \setlength{\parsep}{0pt}  %
    \setlength{\leftmargin}{0pt} %
    \setlength{\labelwidth}{0pt} %
    \setlength{\labelsep}{0pt} %
}}{\end{list}}

\newcommand{\rphi}{r_\Phi}

\title{Harmonic chain with velocity flips: thermalization and kinetic theory}
\author{Jani Lukkarinen\thanks{\emailjani} , Matteo Marcozzi\thanks{\emailmatteo} , Alessia Nota\thanks{\emailalessia} \\[1em]
$\,^*,\,^\dag,\,^\ddag$\UHaddress\\[0.5em]
$\,^\ddag$\UBaddress}
\date{\today}
\newcommand{\email}[1]{E-mail: \tt #1}
\newcommand{\emailjani}{\email{jani.lukkarinen@helsinki.fi}}
\newcommand{\emailmatteo}{\email{matteo.marcozzi@helsinki.fi}}
\newcommand{\emailalessia}{\email{nota@iam.uni-bonn.de}}
\newcommand{\UHaddress}{\em University of Helsinki, Department of Mathematics and Statistics\\
\em P.O. Box 68, FI-00014 Helsingin yliopisto, Finland}
\newcommand{\UBaddress}{\em University of Bonn, Institute for Applied Mathematics\\
\em Endenicher Allee 60, D-53115 Bonn, Germany}

\date{}
\begin{document}
 \selectlanguage{english}
\maketitle

\begin{abstract}
 We consider the detailed structure of correlations in harmonic chains with pinning and a bulk velocity flip noise
 during the heat relaxation phase which occurs on diffusive time scales, for $t=O(L^2)$ where $L$ is the chain length.
 It has been shown earlier that for non-degenerate harmonic interactions these systems thermalize, and the dominant part of the 
 correlations is given by local thermal equilibrium 
 determined by a temperature profile which satisfies a linear heat equation.  
 Here we are concerned with two new aspects about the thermalization process: the first order corrections in $1/L$ to the local
 equilibrium correlations and the applicability of kinetic theory to study the relaxation process.
 Employing previously derived explicit uniform estimates for the temperature profile, we first derive an explicit form for the first order
 corrections to the particle position-momentum correlations.  By suitably revising the definition of the Wigner transform and
 the kinetic scaling limit we derive a phonon Boltzmann equation whose predictions agree with the explicit computation.
 Comparing the two results, the corrections can be understood as arising from two different sources: a current-related term and 
 a correction to the position-position correlations related to spatial changes in the phonon eigenbasis.
 % In other words, standard Fourier's law is satisfied and energy relaxes towards uniform equilibrium state diffusively.
\end{abstract}

\tableofcontents

\section{Introduction}

We consider a harmonic chain with velocity flips, or the \defem{velocity flip model} for short.
The model dynamics consists of a classical Hamiltonian evolution of the particle positions and velocities, as determined by 
a quadratic Hamiltonian, intercepted with random flips of the particle velocities.
This model was first considered in \cite{FFL94}, and it is one of the simplest known particle chain
models which has a finite thermal conductivity and satisfies the time-dependent
Fourier's law \cite{L14,Simon12}.
The model is blessed with many simplifying features which make possible the usually intractable 
rigorous analysis of heat transport properties.  For instance, it is proven in 
\cite{FFL94} that under quite general conditions every translation invariant stationary state of
the infinite chain with a finite entropy density is
given by a mixture of canonical Gibbs states.  This indicates that
temperature is the sole thermodynamic parameter in the velocity flip model 
with pinning.  The numerical 
simulations of the model with boundary heat baths in \cite{DKL11} support these findings and 
provide more information about the resulting nonequilibrium states.
The structure of steady state correlations and energy fluctuations are
discussed in \cite{bkll11b}
with supporting numerical evidence presented in \cite{bkll11}.
The validity of the proposed hydrodynamic limit equations (Fourier's law)
is proven rigorously in \cite{Simon12} (see the Remark after Theorem
1.2.\ for the changes needed in case the model has pinning).

The strategy for proving the hydrodynamic limit in \cite{Simon12} is based on the
relative entropy method introduced by Yau and Varadhan; we refer to \cite{kipnis99} for a review
of the method.  There one studies the relaxation of initial states which are already 
close to a local thermal equilibrium state and as a result one obtains estimates on how local observables,
averaged over regions of size $O(L)$, evolve at diffusive time scales $O(L^2)$.
The method was applied earlier to a similar model with somewhat different stochastic perturbation in \cite{BO05,B07}.
This model shares many features with the velocity flip model with pinning considered here. For instance, also there thermal conductivity 
is constant and hence temperature evolves according to a linear diffusion equation.

A different approach was chosen in \cite{L14} to study
the evolution of the kinetic temperature profile, $T_t(x) = \mean{p_x^2(t)}$, 
where $p_x(t)$ is the momentum  at time $t$ of the particle at the lattice site $x$.
It was first observed that the temperature profile satisfies a closed renewal-type equation,
and the analysis of the properties of the equation lead to a strong, pointwise, control of the errors 
between the temperature profile and its hydrodynamic description by Fourier's law.

The goal of this paper is to clarify the physical meaning of the results in \cite{L14}, 
and to explore its implications on the structure of general local correlations after local equilibrium has been reached.
We consider the evolution of the full spatial covariance matrix of positions and momenta, 
and by defining a suitable Wigner function from the covariance matrix, 
we compute the first order corrections to the local thermal equilibrium.
The first order correction, at diffusive time scales $t=O(L^2)$, turns out to be 
proportional to the temperature gradient, and hence is $O(L^{-1})$.  
In particular, we expect these results to be valid also for the leading covariance 
in a nonequilibrium steady state of the velocity flip model 
induced, for instance, by boundary thermostats.  More precisely, 
we expect that the local correlations sufficiently far away from the boundary
are then given by the appropriate equilibrium correlations with 
the leading correction given by the first order term derived here.

In the first part of the paper, Section \ref{sec:explicitest}, we deal with a periodic chain of length $ L$
under the same assumptions as used in \cite{L14}.  In particular, 
the stochastic flip rate $ \gamma$ is assumed to be sufficiently large compared to the Hamiltonian dispersion relation. 
Then the estimates derived in \cite{L14} for the kinetic temperature profile can be applied to study
the evolution of the full covariance matrix.  This leads to an explicit, fairly simple, form for the 
first order correction, with rigorous upper bounds for the magnitude of the higher order corrections.

The simple form of the first order corrections begs for an explanation.
According to the Fourier's law the energy current is proportional to the temperature gradient, and thus necessarily $O(L^{-1})$, 
and this is indeed the dominant correction found in the position-momentum correlations.  However, 
there are also other corrections of the same order, namely in the position-position correlations, 
while momentum-momentum correlations feature no such corrections.
In the second part, Section \ref{sec:Kinetic}, we derive the same dominant correction term from the kinetic theory of phonons.
This provides a qualitative description of the correction and explains also the origin of the 
position-position correlations.

The kinetic theory of a similar system---merely 
with somewhat different, momentum conserving noise---has been derived in \cite{Basile2009}.  
It is shown there that a kinetic scaling limit of a  lattice Wigner function of phonon modes
satisfies a linear phonon Boltzmann equation, and thus its evolution can be studied via the solutions 
of the Boltzmann equation.  Here we diverge from the standard scheme on two accounts.
Firstly, we employ a somewhat different definition of the Wigner function in which 
explicit real-valuedness is sacrificed for simpler analytical estimates.  Several alternative definitions and basic properties
of more standard Wigner functions for continuum and lattice waves are available in 
\cite{gerard97,mielke05} and in Appendix B of \cite{ls05}.  Secondly, 
we do not take any scaling limits explicitly but rather introduce spatial averaging 
into the definition of Wigner function.  This allows separating 
phonon collisions from the large scale transport without
resorting to scaling limits.

The resulting kinetic theory of the velocity flip model is determined 
by a phonon Boltzmann equation with a very simple collision operator and solving the equation is most standard.
However, proper application of the result for spatially inhomogeneous states requires also 
analysis of polarization effects, in particular, of the evolution of field self-correlations.
Our treatment
of the kinetic theory is not fully rigorous but it is vindicated in the answer to the question
about first order corrections to local equilibrium at diffusive scales: the corrections 
are found to be entirely consistent with the previous rigorously derived result.  
In particular,
the somewhat unexpected position-position correlations are found to arise from  
changes in the phonon eigenbasis resulting from the inhomogeneities in the energy profile. 

We compare the two results in more detail in Section \ref{sec:conclusions}.  The three Appendices contain more details about
some of the main computations used in the text.

Let us emphasize that we only consider models with pinning here.  If the
onsite potential is absent, a second locally conserved field related to the \defem{tension} 
in the chain appears, in addition to the present temperature field.
For results about the hydrodynamics of the velocity flip model without pinning, we refer to \cite{Simon12,bkll11b,bkll11}. 
A more general overview about thermal transport in similar particle chains can be found in \cite{Lepri2016}. 
In particular, in Chapter 5 \cite{bbjko16} the results of \cite{Basile2009} are reviewed along with other rigorous works 
dealing with similar stochastic models.

\section{Evolution of the first two moments in the velocity flip model}
\label{sec:model}

In this section we briefly recall the velocity flip model and the notations used in \cite{L14}. We consider a one-dimensional periodic chain (circle) of $ L$ particles and we parametrize the sites on the chain by
\begin{align}
  & \Lambda_L := \Bigl\{-\frac{L-1}{2},\ldots,\frac{L-1}{2}\Bigr\} \, , \qquad
\text{if $L$ is odd},\\
  &\Lambda_L := \Bigl\{-\frac{L}{2}+1,\ldots,\frac{L}{2}\Bigr\}\, , \qquad
\text{if $L$ is even}.
\end{align}
Then always $|\Lambda_L|=L $ and $\Lambda_L\subset \Lambda_{L'}$ if $L\le L'$.
In addition, 
for odd $L$, we have $\Lambda_L = \defset{n\in \Z }{|n|<\frac{L}{2}}$.  We use
periodic arithmetic on $\Lambda_L$, setting 
$x'+x := (x'+x) \bmod \Lambda_L$ for $x',x\in  \Lambda_L$.  On occasion, we would like to stress the 
use of periodicity in the arithmetic, and we use then the somewhat heavier notation $[x'+x]_L$ for $x'+x$.

%Sometimes we will
%need lattices of several different
%sizes simultaneously, and to stress the length of the cyclic group, we then
%employ the notation $[x'+x]_L$ for $x'+x$.  Also, we use $-x$ to denote
%$[0-x]_L$.

The particles are identical with unit mass and interact via linear forces with a finite range given by the potential $\Phi:\Z \to \R$ which is assumed to be
symmetric, $\Phi(-x)=\Phi(x)$.  The range of $\Phi$ is described by $\rphi$ which we assume to be odd and chosen so that
$\Phi(x)=0$ for all $|x| \ge \rphi/2$.  
%Then the support of $\Phi$ lies in $\Lambda_{\rphi}$.  
Moreover, the forces are assumed to be stable and pinning, i.e., 
the Fourier transform $\widehat\Phi$ is required to be strictly positive. 
The related dispersion relation $\omega:\T \to \R$ is defined as $ \omega:= \sqrt{\widehat \Phi}$,
and it
is then a smooth function on the circle $\T:=\R/\Z$
with $\omega_0 := \min_{k\in \T} \omega(k) > 0 $.
The corresponding periodic interaction matrices $\Phi_L \in
\R^{\Lambda_L\times \Lambda_L}$ on $\Lambda_L$ are defined by
\begin{align}\label{eq:defphiL}
  (\Phi_L)_{x',x} := \Phi([x'-x]_L)\, , \quad \text{for all }x',x\in 
\Lambda_L\, .
\end{align}
This clearly results in a real symmetric matrix.

The discrete
Fourier transform $\mathcal{F}_L$ maps functions
$f:\Lambda_L\to\C$ to $\FT{f} : \Lambda_L^* \to
\C$, where $\Lambda^* := \Lambda_L/L \subset (-\frac{1}{2},\frac{1}{2}]$ is the
dual lattice and for $k\in \Lambda_L^*$ we set
\begin{equation} 
   \FT{f}(k) = \sum_{x\in \Lambda_L} f(x) \textrm{e}^{-\textrm{i} 2\pi k \cdot
x} \, .
   \end{equation} 
% The formula holds in fact for all $k\in \mathbb{Z}/L$, in the sense that the
% right hand side is then equal to $\FT{f}(k \bmod \Lambda_L^*)$, i.e., it
% coincides with the periodic extension of $\FT{f}$.
The inverse transform $\mathcal{F}_L^{-1}:g\mapsto \tilde{g}$ is given by
\begin{equation} 
   \tilde{g}(x) = \int_{\Lambda_L^*} \textrm{d} k\,g(k)
     \textrm{e}^{\textrm{i} 2\pi k \cdot x}\,,
\end{equation}
where we use the convenient shorthand notation
\begin{equation} 
  \int_{\Lambda_L^*} \textrm{d} k\, \cdots =
   \frac{1}{|\Lambda_L|} \sum_{k\in \Lambda_L^*} \cdots \, .
\end{equation}
With the above conventions, for any $L\ge \rphi$ we have
\begin{align}
  (\mathcal{F}_L \Phi_L f)(k) = \omega(k)^2 \FT{f}(k)\,
, \quad \text{for all }k\in  \Lambda_L^*\, ,
\end{align}
i.e., the functional form of the interaction  in the Fourier space does not depend on $L$.

We also use $\delta_L$ to denote  a ``discrete $\delta$-function'' on $\Lambda^*_L$,
defined by 
\begin{equation}
  \delta_{L}(k) = |\Lambda_L| \mathbbm{1}(k = 0) \, , \quad \text{for }k\in 
\Lambda^*_L\, .
\end{equation}
Here, and in the following, $\cf$ 
denotes the generic characteristic function: $\cf(P)=1$ if the condition $P$ is
true, and otherwise $\cf(P)=0$.

The linear forces on the circle are then generated by the Hamiltonian
\begin{align}\label{eq:defHLandGL}
  & H_L(X)  := \sum_{x\in \Lambda_L} \frac{1}{2} (X_x^2)^2 
  + \sum_{x',x\in \Lambda_L} \frac{1}{2} X_{x'}^1 X_{x}^1 \Phi([x'-x]_L)
  = \frac{1}{2} X^T \mathcal{G}_L X\, ,\\
  & \mathcal{G}_L := \begin{pmatrix}
                    \Phi_L & 0 \\ 
                    0 & 1
                  \end{pmatrix} \in \R^{(2 \Lambda_L)\times(2 \Lambda_L)} \, ,
\end{align}
on the phase space $X\in \Omega:= \R^{\Lambda_L}\times \R^{\Lambda_L}$. 
The canonical pair of variables for the site $x$ are
the position $q_x := X^1_x$, and the momentum $p_x := X^2_x$.  
By adding to the Hamiltonian evolution a velocity-flip noise, the system can be identified with a Markov process $X(t)$ and
the process generates a Feller semigroup on the space of observables vanishing
at infinity (see \cite{BO11,Simon12}). For $t> 0$ and any $F$ in the domain of the generator $\pazocal{L}$ of the
Feller process the expectation values of $F(X(t))$ satisfy an evolution
equation
\begin{align}\label{eq:mainevoleq}
  &\partial_t \mean{F(X(t))} = \mean{(\pazocal{L} F)(X(t))},
  %\quad \text{where }\mathcal{L} := \mathcal{A} + \mathcal{S}\, ,
    \\ \intertext{where $\pazocal{L} := \pazocal{A} + \pazocal{S}$, with}
  &\pazocal{A} := 
  \sum_{x\in \Lambda_L} \left(X_x^2 \partial_{X_x^1} - (\Phi_L X^1)_x
\partial_{X_x^2}\right)\, ,\\  
  &(\pazocal{S}F)(X) := \frac{\gamma}{2} \sum_{x_0\in \Lambda_L} \left(
F(S_{x_0} X)- F(X) \right), \quad \gamma>0\, ,\\
  &(S_{x_0} X)_x^i := \begin{cases}
                 -X_x^i\, , & \text{if }i=2 \text{ and }x=x_0\, , \\
                 X_x^i\, , & \text{otherwise}\, .
               \end{cases}
\end{align}

Let $\bar{q}_t=\E[ q_t] $ and $\bar{p}_t=\E[ p_t]$. Then
\begin{align}\label{eq:evol}
&\partial_t \bar{q}_t(x)=\E[ \mathcal{L}q_t(x)]=\bar{p}_t (x)\, , \\\nonumber
&\partial_t \bar{p}_t(x)=\E[ \mathcal{L}p_t(x)]=
%-\langle \left(\Phi_Lq_t\right)(x)\rangle+ \frac{\gamma}{2}\left(\langle -p_t(x)\rangle - \langle p_t(x)\rangle \right) \\\nonumber
% &\qquad \quad = 
-\left(\Phi_L\bar{q}_t\right)(x)- \gamma \;\bar{p}_t (x)\, . \nonumber
\end{align}
We set 
\begin{equation}\label{def:M_gm_lrg}
M_{\gamma}^{}(x,y)=
\begin{pmatrix}
0 & \Phi_L(x-y) \\     -\mathbb{1}(x=y)  & \gamma\, \mathbb{1}(x=y)
\end{pmatrix}
\end{equation}
so that we can rewrite \eqref{eq:evol} in a compact form as
\begin{equation}
\partial_t
\begin{pmatrix}
\bar{q}_t \\   \bar{p}_t
\end{pmatrix}
=
-M_{\gamma}^{\top}
\begin{pmatrix}
\bar{q}_t \\   \bar{p}_t
\end{pmatrix}.
\end{equation}
It follows that
\begin{equation}
\begin{pmatrix}
\bar{q}_t \\   \bar{p}_t
\end{pmatrix}
=
\rme^{-t\,M_{\gamma}^{\top}}
\begin{pmatrix}
\bar{q}_0 \\   \bar{p}_0
\end{pmatrix}.\vspace{2mm}
\end{equation}

The full spacial position-momentum covariance matrix is defined by
\begin{align}\label{eq:full_corr}
C_t^{ij}(x,y):= \E[ X^i_x(t) X^j_y(t)  ] \, .
\end{align}
Strictly speaking, $C_t$ denotes the matrix of second moments, and to get the covariance matrix we should subtract 
the appropriate products of the mean values $\bar{q}_t$ and $\bar{p}_t$.  However, as we will
prove later using the above explicit solutions,
the mean values decay to zero exponentially fast on time scale $O(\gamma^{-1})$, and hence the difference between 
$C_t$ and the covariance matrix is exponentially small in the length $L$ for diffusive time scales $t=O(L^2)$.
Thus the distinction is not relevant in the case considered here.

\begin{remark}
We observe that in this Section we have used mathematically nonstandard, but common in physics, notations for orders of magnitude.
These notations implicitly assume that it has to be known which quantities are large and which small.  Since one of the aims of the present paper is to explore the importance of scaling limits for the validity of kinetic theory, we now state more explicitly what is meant 
by the above notations.  Time and space scales are always assumed to be ``large enough'', so ``$t=O(L^2)$'' actually
means that there is some constant $C>0$ such that $t\ge C L^2$.  In particular, any such $C$ must be
independent from the lattice size $L$, velocity flips and initial data.  However,
any succeeding bounds are allowed to depend on the choice of $C$ and they might blow up as $C\to 0$ or $C\to \infty$.   
If we considered the limit $L\to\infty$,
this could be written using the standard notation as $t^{-1}= O(L^{-2})$.  We will in fact later 
use the notation $O(L^{-2})$ to denote the order of magnitude of many error terms: its precise meaning is to say that there is a constant $C$ as above such that the term is bounded by $C L^{-2}$ for all large enough $L$.
\end{remark}

According to \eqref{eq:mainevoleq}, the entries of $C_t$ satisfy 
\begin{align}\nonumber 
&\partial_t C_t^{11}(x,y)= C_t^{21}(x,y)+C_t^{12}(x,y)\, , \\\nonumber
&\partial_t C_t^{12}(x,y)= C_t^{22}(x,y) - \left(C_t^{11}\Phi_L\right)(x,y)-\gamma\, C_t^{12}(x,y)\, ,\\\nonumber
&\partial_t C_t^{21}(x,y)=C_t^{22}(x,y) - \left(\Phi_L C_t^{11}\right)(x,y)-\gamma\, C_t^{21}(x,y)\, ,\\      \nonumber 
&\partial_t C_t^{22}(x,y)= -\left(\Phi_L C_t^{12}\right)(x,y) - \left(C_t^{21}\Phi_L\right)(x,y)-2\gamma\, C_t^{22}(x,y)+2\gamma \mathbb{1}(x=y)T_t(x)\, . \nonumber
\end{align}
Here $T_t(x)=\E[ p_t(x)^2 ]$ denotes the kinetic temperature at site $x$. 
Therefore, we can write the evolution equation for $C_t$ in a more compact way:
\begin{equation}\label{eq:evolution1}
\partial_t C_t=  -M^{T}_{\gamma}C_t-C_tM_{\gamma}+2\gamma G_t\, ,
 \end{equation}
where
$$G_t(x,y)= 
\begin{pmatrix}
0 & 0 \\     0  &   \mathbb{1}(x=y) T_t(x)
\end{pmatrix}.
$$
%We notice that $G_t$  depends only on the kinetic temperatures $T_t(x)$. %$T_t(x)=\langle p_t(x)^2 \rangle$. 
The matrix evolution equation \eqref{eq:evolution1} can be rewritten in Duhamel's form, so that only the last $G_t$-term remains
as a perturbation.  Namely, integrating the identity
\begin{align}
 \partial_s\left(\rme^{-(t-s) M^{T}_{\gamma}}C_s \rme^{-(t-s) M_{\gamma}}\right)
 = \rme^{-(t-s) M^{T}_{\gamma}}
 \left(M^{T}_{\gamma}C_s+\partial_s C_s +C_s M_{\gamma}\right) \rme^{-(t-s) M_{\gamma}}
\end{align}
over $s$ from $0$ to $t$, we find that any solution to (\ref{eq:evolution1}) also satisfies
\begin{align}\label{eq:evolution}
C_t = \rme^{-t M^{T}_{\gamma}}C_0 \rme^{-t M_{\gamma}}+2\gamma\int_0^t \rmd s\, \rme^{-(t-s) M^{T}_{\gamma}}G_s \rme^{-(t-s) M_{\gamma}}.
  \end{align}
  
In fact, the right hand side in \eqref{eq:evolution}
is a known function which thus determines the evolution of the covariance matrix $C_t$ on the left hand side:
the first term on the right depends only on the initial data covariance $C_0$, and the second term on the matrix $G_s$.
On the other hand, the matrix $G_s$ is a function of the temperature profile $T_s(x)$ only, and its behaviour
has already been solved in \cite{L14}.  As we will show next, 
the strong control derived for the temperature profile in \cite{L14} suffices to determine the local covariances
up to order $O(L^{-2})$ at diffusive time scales.

\section{Uniform estimates in the large flip rate regime} \label{sec:explicitest}

\subsection{The main result}

We first consider a regime in which the flip rate is sufficiently large. More precisely in this Section  we assume that 
$ \gamma > 2 \max_{k \in \T}\omega(k)$. Under this condition several analytical results become available from \cite{L14}. We recall that we want to derive a suitable approximation on the diffusive scale of the full spatial position-momentum covariance matrix.  The structure of the correlations is conveniently studied by introducing the following 
variant of Wigner functions,
\begin{align}\label{eq:U_wigner}
U_t(x,k):=\sum_{y \in \Lambda_L}\rme^{-\ci 2\pi k\cdot y}C_t(x,x+y) \, .
\end{align}
This corresponds to taking a Fourier transform of the covariance matrix with respect to the spatial displacement 
at the point $x$.  Whenever the correlations decay at least square summably, this definition results in a \defem{function}
of $x,k$ instead of a distribution as can occur in other alternatives.  We discuss lattice Wigner functions in more detail
later together with the kinetic theory description in Section \ref{sec:Wignernew}.

To get a more explicit expression for $ U_t$,  we use (\ref{eq:evolution}) and write  
\begin{align}\label{eq:evolution2} 
C_t(x,x+y) &= \sum_{z_1,z_2 \in \Lambda_L}\left(\rme^{-t M^{T}_{\gamma}}\right)_{x,z_1}C_0(z_1,z_2)\left(e^{-t M_{\gamma}}\right)_{z_2,x+y}\\\nonumber & \quad
+2\gamma \int_0^t \rmd s\, \sum_{z_1,z_2 \in \Lambda_L}\left(\rme^{-(t-s) M^{T}_{\gamma}}\right)_{x,z_1}G_s(z_1,z_2) \left(\rme^{-(t-s) M_{\gamma}}\right)_{z_2,x+y}\, .
\end{align}
Thanks to the translation invariance of the matrix $M_\gamma$, we can define a matrix $A_t$ by the condition
$A_t(x-y)=\left(\rme^{-t M_{\gamma}}\right)_{x,y}$.  As shown in \cite[Appendix A]{L14}, its Fourier transform is 
\begin{align}\label{def:A_tk}
\widehat{A}_t(k)&= \sum_{\sigma = \pm 1} \frac{\rme^{-t \mu_{\sigma}(k)}}{\mu_{\sigma}(k)-\mu_{-\sigma}(k)}
\begin{pmatrix}
-\mu_{-\sigma}(k) & \omega(k)^2\\    -1  &  \mu_{\sigma}(k)
\end{pmatrix} \\\nonumber
&= \frac{\rme^{-\gamma t /2}}{\Omega}
\begin{pmatrix}
\frac{\gamma}{2}\sinh \Omega t + \Omega \cosh \Omega t & -\omega(k)^2 \sinh \Omega t \\  \sinh \Omega t  & -\frac{\gamma}{2}\sinh \Omega t + \Omega  \cosh \Omega t
\end{pmatrix} 
\end{align}
with $\Omega = (\gamma /2) \sqrt{1 - (2 \omega(k)/\gamma)^2} < \gamma /2$
and $ \mu_{\sigma}(k)= \gamma /2 +  \sigma \Omega(k)$. (To facilitate comparison, let us point out that the function ``$\Omega$'' was denoted by ``$u$'', and only the second column of $A$ was used in \cite{L14}.)
%and $ \omega(k) = \sqrt{\widehat{\Phi}(k)}$  
Since $ \omega(k)=\omega(-k)$, it follows that $ \widehat{A}_t(k)=\widehat{A}_t(-k)$ and thus also $ A_t(x)=A_t(-x)$ 
because $ A_t$ is real-valued. Let $ P^{(2)}$ denote the projection matrix to the second component, i.e., it is the diagonal $ 2 \times 2$-matrix defined as $P^{(2)}=\diag (0,1)$. Then, thanks to \eqref{eq:U_wigner}, \eqref{eq:evolution2} and \eqref{def:A_tk} we can rewrite $ U_t(x,k)$ as
\begin{align} \label{eq:U_ev0}
U_t(x,k)&= \sum_{y,z \in \Lambda_L}\rme^{-\ci 2\pi k\cdot y}A^{\top}_t(z-x) U_0(z,k) A_t(x+y-z)
\\ \nonumber  &\quad 
+ 2\gamma \int_0^t \rmd s\, \sum_{y,z \in \Lambda_L}\rme^{-\ci 2\pi k\cdot y} T_s(z) A^{\top}_{t-s}(z-x) P^{(2)}A_{t-s}(x+y-z).
 \end{align}
We rename the first term depending on the initial data as 
\begin{align}\label{eq:source_term}
Z_t(x,k): =  \sum_{y,z \in \Lambda_L}\rme^{-\ci 2\pi k\cdot y}A^{\top}_t(z-x) U_0(z,k) A_t(x+y-z)
\end{align}
and, by shifting the summation and integration variables, find
\begin{align}\label{eq:U_ev}
U_t(x,k) 
& = Z_t(x,k)+ 2\gamma \int_0^t \rmd s\, \sum_{y,z \in \Lambda_L}\rme^{-\ci 2\pi k\cdot y} T_{t-s}(x+z) A^{\top}_s(z) P^{(2)} A_s(y-z).
\end{align}

We now define matrices
\begin{align}%\label{eq:Acal}
\mathcal{A}_s(k) & := \widehat{A}^{\top}_s(k) P^{(2)}
\widehat{A}_s(k), \ \ \
\tilde{\mathcal{A}}_s(k)  := \frac{1}{2 \pi}(\partial_k\widehat{A}^{\top}_s(k)) P^{(2)}
\widehat{A}_s(k) 
\end{align}
and
\begin{align*}
{U}_{0}(k) := 2 \gamma \int_0^{\infty} \rmd t \, \mathcal{A}_t(k) = 
 \begin{pmatrix}
 \omega (k)^{-2} & 0 \\
 0 & 1
 \end{pmatrix},\, \ \ \ 
{U}_{1}(k) := 2 \gamma \int_0^{\infty} \rmd t \, \tilde{\mathcal{A}}_t(k)\, .
\end{align*}
Then $ \partial_k {U}_0(k) = 2\pi({U}_1(k) + {U}_1(k) ^{\top})$, and thus we have
\begin{align}%\label{eq:tildeQk}
{U}_1(k) = \frac{1}{2 \pi}
\begin{pmatrix}
-\omega(k)^{-3} \partial_k \omega(k) & q(k) \\
-q(k) & 0
\end{pmatrix}
\end{align}
where 
\begin{equation}\label{eq:qk}
 q(k)=2 \gamma\int_0^{\infty} \rmd t \, \widehat{A}_t(k)^{22} \partial_k \widehat{A}_t(k)^{21}.%=\frac{\partial_k \omega(k)}{ \gamma \omega(k)}\,.
\end{equation}
In Appendix \ref{app} we show by an explicit computation that
\begin{equation}\label{eq:qkexplicit}
 q(k)=\frac{\partial_k \omega(k)}{ \gamma \omega(k)}\,.
\end{equation}
The dispersion relation determines the velocity of the lattice waves with wave number $k$, and with the present choices of normalization, the velocity is given by  $v(k):= \partial_k \omega(k)/2 \pi $.  Hence,
\begin{align}\label{eq:tildeQk}
{U}_1(k) = -\frac{v(k)}{\omega(k)}
\begin{pmatrix}
\omega(k)^{-2} & -\gamma^{-1} \\
\gamma^{-1} & 0
\end{pmatrix}.
\end{align}

We are interested in controlling the behaviour of $ U_t(x,k)$ at the diffusive scale $ t=O(L^2)$.  
We rely on the estimates derived in \cite{L14} and, for the sake of completeness, let us begin by 
summarizing the necessary assumptions from \cite{L14}.  
%We are now ready to state our main rigorous result whose assumptions will be discussed extensively in the proof given in section \ref{proof}.

\begin{assumption}\label{th:phiassump}
  We assume that the map $\Phi:\Z\to \R$ and the flip rate $\gamma$ satisfy all 
  of the following properties where $\omega(k)=\sqrt{\FT{\Phi}(k)}$
  denotes the related phonon dispersion relation:
  \begin{enumerate}
  \setlength{\itemsep}{0pt}
    \item {\em (exponential decay)}\/ There are $C,\delta>0$ such $|\Phi(x)|\le
C \rme^{-\delta |x|}$ for all $x\in \Z$,
    \item {\em (symmetry)}\/ $\Phi(-x)=\Phi(x)$  for all $x\in \Z$,
    \item {\em (pinning)}\/ There is $\omega_0>0$ such that $\omega(k)\ge
\omega_0$ for all $k\in \T$,
    \item {\em (noise dominates)}\/ $\gamma>2 \max_{k\in \T} \omega(k)$,
    \item {\em (harmonic forces are nondegenerate)}\/ For any $\vep>0$ there is
$C_\vep>0$ such that
    \begin{align}\label{eq:nondegcond}
      \int_{0}^\infty\!\rmd t \int_{\T}\! \rmd k \,
\left(F_t\Bigl(k+\frac{k_0}{2}\Bigr)-F_t\Bigl(k-\frac{k_0}{2}
\Bigr)\right)^2 \ge C_\vep\, ,\quad
      \text{whenever } \vep\le |k_0|\le \frac{1}{2}\, ,
    \end{align}
    where $F_t(k):=\widehat{A}^{22}_t(k)$, for $\widehat{A}$ defined in (\ref{def:A_tk}).
  \end{enumerate}
\end{assumption}

These assumptions are satisfied for instance
by the nearest neighbor interactions, for which $\omega(k) =
\sqrt{\omega_0^2+ 4 \sin^2(\pi k)}$, whenever $\omega_0>0$ and $\gamma>2 \sqrt{\omega_0^2+ 4}$. We now state the first result of this paper.

\begin{theorem}\label{theom}
Suppose that Assumption \ref{th:phiassump} holds.  Then there is $L_0>0$ such that 
for any $\pazocal{E}_0,c_0>0$ we can find a constant $c_1>0$ using which 
the following result holds for every $L\ge L_0$.

Assume that the initial state is such that its energy density, 
$\pazocal{E}:= |\Lambda_L|^{-1}\mean{H_L(X(0))}$, is bounded by $\pazocal{E}_0$, that is, 
suppose that $ \pazocal E \leq \pazocal{E}_0$.  
Then  
$ U_t(x,k)$ defined in \eqref{eq:U_wigner} yields a finite function of $x,k$
which satisfies for every $t\ge c_0 L^2$, $x\in \Lambda_L$, $k\in \Lambda^*_L$, $i,j=1,2 $,
\begin{align}\label{eq:thU_gammalarge}
\left|U^{ij}_t(x,k) - \left(T_t(x) U^{ij}_0(k) + \ci \nabla_x T_t(x) U^{ij}_1(k)\right)\right|\le c_1 L^{-2}\, ,
\end{align}
where $\nabla_x T_t(x) := T_t(x+1) -T_t(x) $ denotes a discrete gradient.
\end{theorem}

In the above, the constant $c_1$ may thus depend on $\pazocal{E}_0$, $c_0$, and $L_0$ but it is otherwise
independent of the initial data and of $L$.

The bound in (\ref{eq:thU_gammalarge}) can also be written as
\begin{align}\label{eq:U_gammalarge}
\begin{pmatrix}
U_t^{11}(x,k) & U_t^{12}(x,k) \\
U_t^{21}(x,k) & U_t^{22}(x,k)
\end{pmatrix} = T_t(x)
\begin{pmatrix}
\omega(k)^{-2} & 0 \\
0 & 1
\end{pmatrix} - \frac{\ci v(k) \nabla_x T_t(x)}{\omega(k)} 
\begin{pmatrix}
\omega(k)^{-2} & -\gamma^{-1} \\
\gamma^{-1} & 0
\end{pmatrix} + O(L^{-2}) \, .
\end{align}
Here the $(2,2)$-component of the dominant first term on the right hand side 
corresponds to the diffusive temperature profile found already in \cite{L14}.
Together with the other three matrix components, the first term gives the expected
local thermal equilibrium correlations since the $(q,p)$-correlation matrix of the equilibrium Gibbs
state at temperature $T$ is equal to
\[
T \begin{pmatrix}
 \Phi^{-1} & 0 \\
0 & 1
\end{pmatrix} \, .
\]
The second term 
on the right hand side is an $O(L^{-1})$ correction to the local equilibrium correlations.
Its off-diagonal components can be interpreted as ``current terms'' while
the origin of the diagonal terms will be 
clarified by the kinetic theory description discussed in Section \ref{sec:conclusions}.

\subsection{Proof of Theorem \ref{theom}}\label{proof}

All computations in this subsection are made supposing that Assumption \ref{th:phiassump} holds.  Since then also  Assumptions 4.1 and 4.3 in \cite{L14} are valid, this will allow directly applying all results derived in that reference.

Let us begin the proof of the theorem by explaining how the assumptions, in particular the boundedness of the initial energy density, immediately yield an upper bound for 
the function $U_t$ proving, in particular, that it is finite.  First, by the discrete Plancherel theorem and using the assumed pinning property,
we have for any real $q$
\begin{align}
 \omega_0^2 \sum_{x\in \Lambda_L} q_x^2 =\omega_0^2 \int_{\Lambda_L^*} \textrm{d} k\,|\FT{q}(k)|^2
 \le \int_{\Lambda_L^*} \textrm{d} k\,\omega(k)^2 |\FT{q}(k)|^2 = \sum_{x,y\in \Lambda_L} (\Phi_L)_{x,y} q_x q_y\,.
\end{align}
Therefore, for any real $X=(q,p)$ we have
\begin{align}
 \sum_{x\in \Lambda_L} p_x^2\le 2 H_L(X)\qand \sum_{x\in \Lambda_L} q_x^2 \le 2 \omega_0^{-2} H_L(X)\, .
\end{align}
Thus the assumption $ \pazocal E \leq \pazocal{E}_0<\infty$ and the conservation of the total energy imply that 
for $i=1,2$
\begin{align}
 \sum_{x\in \Lambda_L}\E[(X^i_x(t))^2]\le 2(1+\omega_0^{-2}) \E[H_L(X(t))] = 
2(1+\omega_0^{-2}) \E[H_L(X(0))] =
 2 L(1+\omega_0^{-2})\pazocal{E}<\infty \, .
\end{align}
Therefore, by the Schwarz inequality,
\begin{align}\label{eq:Uapbound}
& |U^{ij}_t(x,k)|\le \sum_{y \in \Lambda_L} \E\left[|X^i_x(t)||X^j_{x+y}(t)|\right] 
\le (L \E[|X^i_x(t)|^2])^{1/2}
\left(\E\!\left[\sum_{x' \in \Lambda_L} |X^j_{x'}(t)|^2\right]\right)^{1/2} \nonumber \\
& \quad 
\le 2(1+\omega_0^{-2}) \pazocal{E} L^{3/2} < \infty
\, .
\end{align}
Since $ \pazocal E \leq \pazocal{E}_0$, this shows that $U_t$ is finite and $O(L^{3/2})$.  The theorem
significantly improves this a priori bound for diffusive times since it implies that then $U_t=O(1)$.

\subsubsection{Review of the properties of $ T_t(x)$}

In this section we collect from \cite{L14} all the necessary ingredients for the derivation of equation \eqref{eq:thU_gammalarge}. We will adopt the notation $a \lesssim b $ to indicate $ a \leq C b$, where $ C$ is a constant which might depend on $ \gamma$ and 
the function $\omega$, but not on $ L$, $ t$ or the initial data. 
Furthermore, we will use for matrices the elementwise sum norm defined as
\begin{align}\label{eq:matrix_norm}
\norm{B} := \sum^{n}_{i,j=1} | B_{ij}|\, ,
\end{align}
where $ B \in \C^{n \times n}$. 
All finite matrix norms are equivalent, but the above choice is convenient for our purposes, in particular, since it is 
sub-multiplicative, i.e., always $ \norm{A B} \leq \norm{A} \norm{B}$.

Thanks to Lemma 4.6 in \cite{L14} we know that there exist strictly positive constants $\gamma_2$ and $\delta_0 $ such that
\begin{enumerate}
\item The entries of the matrix $\widehat{A}_t(k)$ belong to $ C^1([0,\infty)\times \T)$.
\item For every $k\in \T$ and $t\geq 0 $
\begin{align}\label{eq:Ak_estimates}
\|\widehat{A}_t(k)\|\lesssim \,\rme^{-\delta_0t}, \ \ \ \|\partial_t\widehat{A}_t(k)\| \lesssim \, \rme^{-\delta_0t}, \ \ \ \|\partial_k\widehat{A}_t(k)\| \lesssim \, \rme^{-\delta_0t/2} .
\end{align}
%where $ 0 < \delta_0 \leq \Omega(k)-\gamma /2$.
% \item $-\widehat{A}_t^{i}(k)\geq c_1 \rme^{-\gamma t/2}$ for all $t\geq t_1$ and $k\in\T$.
\item For all $x\in\Z$ and $t\geq 0$  
%The functions ${A}_{t}^{2i}(x):=\int_{\Lambda_L^*} \rmd k \,\rme^{\ci 2\pi x\cdot k}\widehat{A}_t^{2i}(k),$ $x\in \Z,$ satisfy 
\begin{align}\label{eq:Ax_estimates}
\|{A}_{t}(x)\|\lesssim \, \rme^{-\delta_0t/2-\gamma_2|x|}.
\end{align}
\end{enumerate}
To be more precise, the bounds \eqref{eq:Ak_estimates} and \eqref{eq:Ax_estimates} are only proven 
for the absolute value of the entries $ \widehat A_t^{i2}(k)$ and $ A_t^{i2}(x)$ for $ i=1,2$ in Lemma 4.6 of \cite{L14}.  However, as is evident from 
\eqref{def:A_tk}, all the entries of $\widehat A_t(k) $ have the same analyticity and decay properties. Thus, these
derivations 
can be extended directly to every matrix element, and hence also to the matrix norm \eqref{eq:matrix_norm}.
Note that from \eqref{eq:Ak_estimates} we immediately obtain
\begin{align}\label{eq:Acal}
\norm{\mathcal{A}_t} \lesssim \,  \rme^{-\delta_0 t}, \qquad\norm{\tilde{\mathcal{A}}_t} \lesssim \,  \rme^{-\delta_0 t}.
\end{align}

% To obtain a  uniform control of the temperature profile, we need an extra assumption on the harmonic interactions to exclude certain degenerate lattice structures, such as 
% splitting of the chain into mutually noninteracting components.  A sufficient condition is given in Eq.~(4.8)
% of \cite{L14}, but the precise form is not important here; let us merely mention that 
% for instance the nearest neighbour interactions with pinning satisfy all of the requirements.

A renewal equation was derived in \cite{L14}
for the noise-averaged temperature profile $T_t(x)$ and its solution was shown to satisfy a linear diffusion equation
at diffusive time-scales.  Indeed, the defining equation for $T_t(x)$, equation (4.2) in \cite{L14},
is equal to the $((2,x),(2,x))$-component of the Duhamel formula in (\ref{eq:evolution}) and thus their solutions coincide.
Since now Assumptions 4.1 and 4.3 in \cite{L14} hold, and we have also proven that all second moments of $X(0)$ are finite, we can apply Theorem 4.4 in \cite{L14}.  We can thus conclude that there is $L_0>0$ such that for all $L\ge L_0$, $t>0$ and $x\in \Lambda_L$, the temperature profile
$T_t(x) = \mean{p_x^2(t)}$ satisfies

\begin{equation}\label{eq:th1}
\left| T_{t}(x)-(\rme^{-tD }\tau)_x  \right| \lesssim \, \pazocal{E} L  t^{-3/2}
\end{equation}
where the discrete diffusion operator $ D$ is defined by 
$$
(D\phi)_x := \sum_{y \in \Lambda_L} \tilde{K}_y (2 \phi_x - \phi_{x+y} - \phi_{x-y})\, ,
$$
with\footnote{To avoid possible confusion with the particle momenta, we deviate here from the notations in \cite{L14} where ``$K_{t,x}$'' and ``$\tilde{K}_x$'' are denoted by ``$p_{t,x}$'' and ``$\tilde{p}_x$'', respectively.}
\begin{equation}\label{eq:defp}
\tilde{K}_x:=\frac{\gamma}{2}\int_0^\infty \rmd s\, K_{s,x}, \quad K_{t,x}:= 2 \gamma ((\rme^{-t M_{\gamma}})^{22}_{0,x})^2\, .
\end{equation}
The initial data vector $\tau$ for the discrete diffusion has an explicit, but somewhat involved,
dependence on the initial data of the particle system.  Namely, 
\begin{equation}\label{eq:deftau}
\tau_x:= \sum_{y\in  \Lambda_L }\int_{\Lambda_L^{*}}\! \rmd k \,\rme^{2 \pi i k \cdot (x-y)} a(k) \int_0^{\infty} \rmd s \, g_{s,y}\, , 
\end{equation}
where
\begin{align*}
g_{t,x}= \big( \rme^{-tM_{\gamma}^{\top}}\Gamma_x \rme^{-tM_{\gamma}}\big)^{22}(0,0), \quad (\Gamma_x)^{ij}(y,y') := C_0^{ij}(x+y,x+y')\, ,
\end{align*}
and $a(k)$ are explicit constants satisfying $0\le a(k)\lesssim 1$.  
It is proven in Proposition 4.8 of \cite{L14} that $g_{t,x}$ are positive and satisfy a bound
$\sum_x g_{t,x}\lesssim \pazocal{E} L \rme^{-\delta_0 t}$.  Thus the initial data vector $\tau$
and its discrete Fourier transform are bounded by the total energy,
\begin{align}\label{eq:esttau}
|\tau_x | \lesssim \, \pazocal{E} L\,, \quad
|\widehat{\tau}(k) | \lesssim \, \pazocal{E} L\,.
\end{align}

The Fourier transform of the diffusion operator is given by 
\begin{align}%\label{eq:D_est}
\widehat{D}(k)= \sum_{y \in \Lambda_L}(1-\cos(2 \pi k \cdot y)) 2 \tilde{K}_y \, .
\end{align}
It is bounded from both above and below, $0\le \widehat{D}(k)\le 2\gamma$, and 
the assumptions can be used to show that 
its small $k$ behaviour is controlled by the estimates
\begin{align}\label{eq:D_est}
C_1 \min(|k|, \varepsilon_0)^2\le \widehat{D}(k) \leq C_2 k^2\, ,
\end{align}
where $C_1,C_2,\varepsilon_0>0$ are constants of the kind mentioned in the beginning of this section.
The lattice diffusion approximation, $(\rme^{-tD }\tau)_x$, is equal to 
$\int_{\Lambda_L^{*}}\! \rmd k \, \rme^{2 \pi \ci k \cdot x} \rme^{-t \widehat{D}(k)} \FT{\tau}(k)$, and thus
it is bounded by 
\begin{align}%\label{eq:D_est}
|(\rme^{-tD }\tau)_x|\lesssim  \, \pazocal{E} L\, .
\end{align}

Therefore, thanks to \eqref{eq:th1}, for all $ t>0$ and $ x \in \Lambda_L$ we can write 
\begin{align}\label{eq:T_decomp}
T_{t}(x) =(\rme^{-tD} \tau)_x + \delta_t(x)
\end{align}
where for large $ t$ 
\begin{align}\label{eq:delta_tlarge}
|\delta_t(x)| \lesssim \, \pazocal{E} L t^{-3/2}\, .
\end{align}
Since $T_t(x)=\mean{p_x(t)^2}$ is obviously bounded by the total energy, 
we can also conclude validity of the following {\it a priori} bounds
\begin{align}\label{eq:delta_apriori}
|T_t(x)| \lesssim \, \pazocal{E} L\, , \quad  |\delta_t(x)| \lesssim \, \pazocal{E}L\, .
\end{align}
These trivial bounds are used later only to control small values of $t$ for which the more accurate estimate in  \eqref{eq:delta_tlarge} becomes uninformative.

\subsubsection{Derivation of equation \eqref{eq:thU_gammalarge}}\label{sec:derivationeqU_gammalarge}

Now we have all the necessary ingredients to find an approximate evolution equation for the observable $ U_t(x,k)$ at the diffusive scale.  To this end, we now assume that $c_0>0$ is fixed and the initial data satisfies $\pazocal E\le \pazocal E_0$, and we then consider arbitrary $L\ge L_0$ and $t\ge c_0 L^2$.

Let us start by examining the source term in  \eqref{eq:source_term}: by shifting the summation variables we find
\begin{align*}
Z_t(x,k) = & \sum_{y,z \in \Lambda_L}\rme^{-\ci 2\pi k\cdot y}A^{\top}_t(z-x)U_0(z,k)A_t(x+y-z) \\\nonumber
% = & \sum_{y,z'}\rme^{-\ci 2\pi k\cdot y}A^{\top}_t(z')U_0(x+z',k)A_t(y-z') \\\nonumber
= & \sum_{z \in \Lambda_L}\rme^{-\ci 2\pi k\cdot z}A^{\top}_t(z)U_0(x+z,k)\widehat{A}_t(k)\, .
\end{align*}
As proven in (\ref{eq:Uapbound}), the assumptions imply that $\norm{U_{0}(x,k)} \lesssim \pazocal E L^{3/2}$. Then by \eqref{eq:Ax_estimates} and \eqref{eq:Ak_estimates} we get
\begin{align}
\norm{Z_t(x,k)} \leq \sum_{z \in \Lambda_L}\norm{A^{\top}_t(z)} \norm{U_0(x+z,k)} \norm{\widehat{A}_t(k)} 
\lesssim \pazocal E L^{3/2} \rme^{-3\delta_0 t/2} \sum_{z \in \Lambda_L} \rme^{-\gamma_2 |z|} \lesssim \, \pazocal E L^{3/2}\rme^{-3\delta_0 t/2}\, ,
\end{align}
which is exponentially small in $L$ for $t\ge c_0 L^2$ and $\pazocal E\le \pazocal E_0$.
Let us denote terms which are exponentially small in $L$ by $O(\rme^{-\delta L})$ in the following without specifying the exact value of $\delta>0$.   In particular, the value of $\delta$ might vary from one equation to the next.

Hence, we may now conclude that $\norm{Z_t(x,k)} = O(\rme^{-\delta L})$, i.e., that $\norm{Z_t(x,k)} \le c \rme^{-\delta L}$
with a constant $c$ which might depend on $L_0$, $c_0$, and $\pazocal E_0$ but is independent from initial data, $x$, $t$, and $L$.

In order to analyse the second term in \eqref{eq:U_ev}, let us decompose $ T_{t-s}(x+z)$ by \eqref{eq:T_decomp} as 
\begin{align}\label{eq:T_telescope}\nonumber
T_{t-s}(x+z) & =  T_t(x+z) + [T_{t-s}(x+z)-T_t(x+z)]  \\\nonumber
& =  T_t(x+z)  + [(\rme^{-(t-s)D}\tau)_{x+z} - (\rme^{-tD}\tau)_{x+z}]+[\delta_{t-s}(x + z) - \delta_{t}(x+z)]\\\nonumber
& =  T_t(x+z) -  \int_{t-s}^t \rmd {s'} \partial_{s'} (\rme^{-s'D}\tau)_{x+z} +[\delta_{t-s}(x + z) - \delta_{t}(x+z)]  \\\nonumber
& =  T_t(x) +  \int_{t-s}^t \rmd {s'} \int_{\Lambda_L^*} \rmd q \,\rme^{2 \pi \ci q (x+z)}\widehat D(q)\rme^{-s'\widehat D(q)}\widehat \tau (q)   
\\ & \quad  +[\delta_{t-s}(x + z) - \delta_{t}(x+z)]+[T_{t}(x+z)-T_t(x)]
\end{align}
 where in the last passage we wrote $ (\rme^{-s'D}\tau)_{x+z}$ in terms of its Fourier transform. 
 Therefore, from \eqref{eq:U_ev} and \eqref{eq:T_telescope} we get
\begin{align}\label{eq:U1}
U_t(x,k)
&= Z_t(x,k) +  2\gamma T_{t}(x) \int_0^t \rmd s\, \sum_{y,z \in \Lambda_L}\rme^{-\ci 2\pi k\cdot y}A^{\top}_s(z)P^{(2)} A_s(y-z) \\\nonumber
& \quad +  2\gamma \int_0^t \rmd s\, \sum_{y,z \in \Lambda_L}\rme^{-\ci 2\pi k\cdot y}(T_{t-s}(x+z)-T_t(x))A^{\top}_s(z) P^{(2)} A_s(y-z) \\\nonumber
&= Z_t(x,k) +  2\gamma T_t(x)\int_0^{\infty} \rmd s\,  \mathcal{A}_s(k)  + I^{(1)}(k) + I^{(2)}(k) + I^{(3)}(k) + I^{(4)}(k) 
\end{align}
where
\begin{align*}
 I_t^{(1)}(x,k) & =-2\gamma T_t(x)\int_t^{\infty} \rmd s\, \mathcal{A}_s(k) \\\nonumber
 I_t^{(2)}(x,k) & = 2\gamma \int_0^t \rmd s\, \sum_{y,z \in \Lambda_L}\rme^{-\ci 2\pi k\cdot y}
 A^{\top}_s(z) P^{(2)} A_s(y-z)\bigg(\int_{t-s}^t \rmd {s'} \int_{\Lambda_L^*} \rmd q \,\rme^{2 \pi \ci q (x+z)}\widehat D(q)\rme^{-s'\widehat D(q)}\widehat \tau (q) \bigg)
\\\nonumber
& = 2\gamma \int_0^t \rmd s\, \int_{\Lambda_L^*} \rmd q \, \rme^{2 \pi \ci q x}
\widehat A^{\top}_s(k-q) P^{(2)} \widehat A_s(k) \bigg(\int_{t-s}^t \rmd {s'} \widehat D(q)\rme^{-s'\widehat D(q)}\widehat \tau (q) \bigg)\\\nonumber
I_t^{(3)}(x,k) &= 2\gamma \int_0^t \rmd s\, \sum_{z \in \Lambda_L}\rme^{-\ci 2\pi k\cdot z}[\delta_{t-s}(x + z) - \delta_{t}(x+z)]  A^{\top}_s(z) P^{(2)}\widehat A_s(k)
 \\\nonumber
 I_t^{(4)}(x,k) & = 2\gamma \int_0^t \rmd s\, \sum_{z \in \Lambda_L}\rme^{-\ci 2\pi k\cdot z}[T_{t}(x+z)-T_t(x)]  A^{\top}_s(z) P^{(2)} \widehat A_s(k). 
\end{align*}

We now consider separately the terms $I_t^{(1)},$ $I_t^{(2)},$ $I_t^{(3)}$ and $I_t^{(4)}$. The bounds \eqref{eq:Ak_estimates}, \eqref{eq:Ax_estimates}, \eqref{eq:Acal}, \eqref{eq:delta_tlarge} and \eqref{eq:delta_apriori} yield
\begin{align*}
\norm{I^{(1)}_t(x,k)} \leq 2 \gamma |T_t(x) | \int_t^{\infty} \rmd s \norm{\mathcal{A}_s(k)} \lesssim \, \pazocal{E} L \rme^{-\delta_0 t}
\end{align*}
%while for $ I^{(3)}_t(x,k)$ we have
and
\begin{align*}
\norm{I^{(3)}_t(x,k)} & \leq 2\gamma \int_0^t \rmd s\, \sum_{z \in \Lambda_L}|\delta_{t-s}(x + z) - \delta_{t}(x+z)| \norm{ A^{\top}_s(z)} \norm{\widehat A_s(k)} \\\nonumber 
& \lesssim \, \int_0^t \rmd s\, \sum_{z \in \Lambda_L}|\delta_{t-s}(x + z) - \delta_{t}(x+z)| \rme^{-\delta_0 s - \gamma_2 |z|} \\\nonumber
& \lesssim \, \sum_{z \in \Lambda_L} \bigg[ \int_0^{t/2} \rmd s\, |\delta_{t-s}(x + z) - \delta_{t}(x+z)| \rme^{-\delta_0 s - \gamma_2 |z|} \\\nonumber
& \qquad + \int_{t/2}^t \rmd s\, |\delta_{t-s}(x + z) - \delta_{t}(x+z)| \rme^{-\delta_0 s - \gamma_2 |z|} \bigg] \\\nonumber
& \lesssim \,  \pazocal{E} L \bigg[ \int_0^{t/2}\rmd s\, (t-s)^{-3/2} \rme^{-\delta_0 s } + \int_{t/2}^t \rmd s\,\rme^{-\delta_0 s} \bigg] \lesssim \,  \pazocal{E} L t^{-3/2}
\end{align*}
The estimate for $ I_t^{(2)}(x,k)$ is slightly more complicated: thanks 
to \eqref{eq:Ak_estimates}, \eqref{eq:Ax_estimates},  \eqref{eq:D_est} and \eqref{eq:esttau} we get
\begin{align*}
\norm{I_t^{(2)}(x,k)} & \lesssim \,  \pazocal{E} L \int_0^t \rmd s\, \rme^{-\delta_0 s} \int_{t-s}^t \rmd s' \int_{\Lambda_L^*} \rmd q \, \rme^{-s'\widehat{D}(q)}|\widehat{D}(q)|\\\nonumber
% & \lesssim \,  \pazocal{E} L  \int_0^t \rmd s\, \rme^{-\delta_0 s} \int_{t-s}^t \rmd s' \int_{| q| \leq \varepsilon_0} \rmd q \, \rme^{-s'\widehat{D}(q)}(|\widehat{D}(q)-b(q)| + | b(q)|) \\\nonumber
% & +  \pazocal{E} L  \int_0^t \rmd s\, \rme^{-\delta_0 s} \int_{t-s}^t \rmd s' \int_{| q| > \varepsilon_0} \rmd q \, \rme^{-s'\widehat{D}(q)}|\widehat{D}(q)| \\\nonumber
& \lesssim \,  \pazocal{E} L \bigg[  \int_0^t \rmd s\, \rme^{-\delta_0 s} \int_{t-s}^t \rmd s' \int_{| q| \leq \varepsilon_0} \rmd q \, \rme^{-c_1 s' q^2 }q^2 + \int_0^t \rmd s\, \rme^{-\delta_0 s} \int_{t-s}^t \rmd s' \int_{| q| > \varepsilon_0} \rmd q \, \rme^{-s'c_1\varepsilon_0^2} \bigg] \\\nonumber
% & \lesssim \,  \pazocal{E} L  \bigg[  \int_0^t \rmd s\, \rme^{-\delta_0 s} \int_{t-s}^t \rmd s' \int_{| q| \leq \varepsilon_0} \rmd q \, \rme^{-c_1 s' q^2 }q^2 + \int_0^t \rmd s\, \rme^{-\delta_0 s} \int_{t-s}^t \rmd s' \, \rme^{-s'c_1'\varepsilon_0^2} \bigg] \\\nonumber
& \lesssim \,  \pazocal{E} L  \bigg[  \int_0^{t/2} \rmd s\, \rme^{-\delta_0 s} \int_{t-s}^t \rmd s' \int_{| q| \leq \varepsilon_0} \rmd q \, \rme^{-c_1 s' q^2 } q^2  +  \int_{t/2}^t \rmd s\, \rme^{-\delta_0 s} \int_{t-s}^t \rmd s' \, \rme^{-c_1 s' q^2 } q^2 \bigg] \\\nonumber
& \quad +  \pazocal{E} L  \bigg[  \int_0^{t/2} \rmd s\, \rme^{-\delta_0 s} \int_{t-s}^t \rmd s'  \, \rme^{-s'c_1\varepsilon_0^2} + \int_{t/2}^t \rmd s\, \rme^{-\delta_0 s} \int_{t-s}^t \rmd s'  \, \rme^{-s'c_1\varepsilon_0^2} \bigg] \\\nonumber
& =: J_a + J_b + J_c +J_d.
\end{align*}
We now study each $J_i$'s contribution separately: by Lemma 4.11 of \cite{L14}, $\int_{| q| \leq \varepsilon_0} \rmd q \, \rme^{-c_1 s' q^2 } q^2\le 4 (c_1 s')^{-3/2}$, and thus
\begin{align*}
J_a & 
 \lesssim \,  \pazocal{E} L  \int_0^{t/2} \rmd s\, \rme^{-\delta_0 s} \int_{t-s}^t \rmd s'( s')^{-3/2}  \lesssim \,  \pazocal{E} L \int_0^{t/2} \rmd s\, \rme^{-\delta_0 s} s (t- s)^{-3/2} \\\nonumber
& \lesssim \,  \pazocal{E} L t^{-3/2}\int_0^{t/2} \rmd s\, \rme^{-\delta_0 s} s  
 \lesssim \,  \pazocal{E} L t^{-3/2} \\\nonumber 
J_b & 
% = C  \pazocal{E} L \int_{t/2}^t \rmd s\, \rme^{-\delta_0 s} \int_{t-s}^t \rmd s' \int_{| q| \leq \varepsilon_0} \rmd q \, \rme^{-c_1 s' q^2 } q^2 
\lesssim \,  \pazocal{E} L  \int_{t/2}^t \rmd s\, \rme^{-\delta_0 s} s \lesssim \,  \pazocal{E} L  t \rme^{-\delta_0 t/2} \\\nonumber
J_c & 
% = C  \pazocal{E} L \int_0^{t/2} \rmd s\, \rme^{-\delta_0 s} \int_{t-s}^t \rmd s' \int_{| q| > \varepsilon_0} \rmd q \, \rme^{-s'\varepsilon_0^2}  
% \lesssim \,  \pazocal{E} L  \int_0^{t/2} \rmd s\, \rme^{-\delta_0 s}s \, \rme^{-(t-s)c_1'\varepsilon_0^2} 
 \lesssim \,  \pazocal{E} L \int_0^{t/2} \rmd s\, \rme^{-(t-s)c_1 \varepsilon_0^2} s  \lesssim \,  \pazocal{E} L   t\rme^{-t c_1 \varepsilon_0^2/2} \\\nonumber
J_d & 
% = C  \pazocal{E} L \int_{t/2}^t \rmd s\, \rme^{-\delta_0 s} \int_{t-s}^t \rmd s'  \, \rme^{-s'c_1'\varepsilon_0^2}  
\lesssim \,  \pazocal{E} L  \int_{t/2}^t \rmd s\, \rme^{-\delta_0 s} s \lesssim \,  \pazocal{E} L t  \rme^{-\delta_0 t/2 }. 
\end{align*}
From the computations above it follows that, on the diffusive scale $t\ge c_0 L^2$, the sum of the first three contributions is $O(L^{-2})$, i.e., $I^{(1)} + I^{(2)}+ I^{(3)} = O(L^{-2})$.

We now focus on $ I^{(4)}_t(x,k)$ from which the dominant correction arises. We define the discrete gradient as $(\nabla f)(x):=f(x+1)-f(x) $. If $ y \geq 0$, by induction one can check that
\begin{align}\label{eq:discr_der}
f(x+y)-f(x)= y (\nabla f)(x) + \sum_{z=0}^{y-1} [(\nabla f)(x+z) - (\nabla f)(x)]
\end{align}
and, if $ y <0$, by using \eqref{eq:discr_der}, one gets
\begin{align}\nonumber
& f(x+y)-f(x) \\ 
& = y (\nabla f)(x) + y[(\nabla f)(x+y) -(\nabla f)(x)]  - \sum_{z=0}^{|y|-1} [(\nabla f)(x+y + z) - (\nabla f)(x + y)].
\label{eq:discr_der2}
\end{align}
For any $ z \in \Lambda_L$, let us define 
\begin{align}\label{eq:R_corr}
\pazocal{R}(f;x,y) := f(x+y)-f(x)- y (\nabla f)(x)
\end{align}
which is the correction to the first order discrete Taylor expansion of $ f(x+y)$ around $ x$. Then, given the Fourier transform $ \widehat{f} = \mathcal{F} f$, by exploiting \eqref{eq:discr_der}, \eqref{eq:discr_der2} and the inequality $ |\rme^{\ci r} - 1-\ci r| \leq r^2/2$, valid for $ r \in \R$, one has
\begin{align}\label{eq:second_discr_der}
|\pazocal{R}(f;x,y)| \lesssim \, y^2 \int_{\Lambda^*_L} \rmd q \, q^2 |\widehat{f} (q) |.
\end{align}
On the other hand, we also have the trivial bound
\begin{align}\label{eq:trivial_R}
|\pazocal{R}(f;x,y)| \lesssim \, |y| \sup_{x \in \Lambda_L} |f(x)|.
\end{align}

Thus, by \eqref{eq:R_corr} we can split $ I^{(4)}_t(x,k)$ as follows
\begin{align}\label{eq:I4}\nonumber
I^{(4)}_t(x,k) & =  2\gamma \int_0^t \rmd s\, \sum_{z \in \Lambda_L}\rme^{-\ci 2\pi k\cdot z}[T_{t}(x+z)-T_t(x)]  A^{\top}_s(z) P^{(2)} \widehat A_s(k) \\\nonumber
& =  2\gamma (\nabla T_t)(x)\int_0^t \rmd s\, \sum_{z \in \Lambda_L}\rme^{-\ci 2\pi k\cdot z} z A^{\top}_s(z) P^{(2)} \widehat A_s(k) \\\nonumber 
& \quad +2\gamma \int_0^t \rmd s\, \sum_{z \in \Lambda_L}\rme^{-\ci 2\pi k\cdot z}\pazocal{R}(T_t;x,z)  A^{\top}_s(z) P^{(2)} \widehat A_s(k) \\\nonumber
& =  \frac{ 2\ci\gamma}{2 \pi } \nabla T_{t}(x)\int_0^{\infty} \rmd s\, (\partial_k \widehat{A}_s^{\top}(k)) 
P^{(2)} \widehat{A}_s(k)- \frac{ 2\ci\gamma}{2 \pi } \nabla T_{t}(x)\int_t^{\infty} \rmd s\, (\partial_k \widehat{A}_s^{\top}(k)) 
P^{(2)} \widehat{A}_s(k) \\
& \quad +2\gamma \int_0^t \rmd s\, \sum_{z \in \Lambda_L}\rme^{-\ci 2\pi k\cdot z}\pazocal{R}(T_t;x,z)  A^{\top}_s(z) P^{(2)} \widehat A_s(k) 
+ O(\rme^{-\delta L})\, .
\end{align}

The additional exponentially small correction $O(\rme^{-\delta L})$ arises from the following mismatch between discrete Fourier transform and the Fourier series.  Suppose $f:\T^d\to \C$ is a continuously differentiable function whose Fourier coefficients are exponentially decaying, i.e., 
$F(n):=\int_{\T}\!\rmd k\, f(k) \rme^{\ci 2\pi n\cdot k}=O(\rme^{-\delta |n|})$.  Then the Fourier series of $F$
converges at every point to $f$, i.e., pointwise $f(k)=\sum_{n\in \Z^d} \rme^{-\ci 2\pi n \cdot k}F(n)$, $k\in \T^d$.
Thus the discrete Fourier transform of $f$ restricted to $\Lambda_L^*$ is equal to  
$\tilde{f}(x)= \int_{\Lambda_L^*} \rmd q \, \rme^{2 \pi \ci q \cdot x} f(q)=
\sum_{m\in \Z^d} F(x+L m)$, for all $x\in \Lambda_L$.
Moreover, for any $k\in \Lambda_L^*$,
\begin{align}\label{eq:discretederivcorrect}
 & \sum_{x \in \Lambda_L} x\rme^{-\ci 2\pi k\cdot x} \tilde{f}(x) =
 \sum_{m\in \Z^d} 
 \sum_{x \in \Lambda_L} x\rme^{-\ci 2\pi k\cdot x} F(x+L m)  \\\nonumber
& \quad
 = 
 \sum_{m\in \Z^d} 
 \sum_{x \in \Lambda_L} (x+L m-Lm)\rme^{-\ci 2\pi k\cdot (x+L m)} F(x+L m)
 \\\nonumber
& \quad
 = 
 \sum_{y\in \Z^d} y \rme^{-\ci 2\pi k\cdot y} F(y)- 
 \sum_{m\in \Z^d,m\ne 0} L m
 \sum_{x \in \Lambda_L} \rme^{-\ci 2\pi k\cdot (x+L m)} F(x+L m)
 \\\nonumber
& \quad
 = \frac{\ci}{2\pi}\nabla f(k) + O(\rme^{-\delta L/4})\, ,
\end{align}
where $\nabla f$ denotes the ordinary (continuum) gradient of $f$.

We identify the first term on the right hand side of \eqref{eq:I4} as the claimed correction term, more precisely 
\begin{align}
\mathcal{J}_t(x,k)=\frac{ 2\gamma}{2 \pi } \nabla T_{t}(x)\int_0^{\infty} \rmd s\, (\partial_k \widehat{A}_s^{\top}(k)) 
P^{(2)} \widehat{A}_s(k) = 2\gamma \nabla T_{t}(x)\int_0^{\infty} \rmd s\, \tilde{ \mathcal{A}}_s(k).
\end{align}
We are left with showing that the second and third term on the right hand side of \eqref{eq:I4} are $ O(L^{-2})$ at the diffusive scale.  By \eqref{eq:delta_apriori} and \eqref{eq:Acal} for the second one we simply have
\begin{align}
\bigg \| \frac{ 2\ci\gamma}{2 \pi } \nabla T_{t}(x)\int_t^{\infty} \rmd s\, (\partial_k \widehat{A}_s^{\top}(k)) 
P^{(2)} \widehat{A}_s(k) \bigg \| \lesssim \,  \pazocal{E} L \rme^{-\delta_0 t}.
\end{align}
By using \eqref{eq:T_decomp} we can decompose the third term on the right hand side of \eqref{eq:I4} as 
\begin{align*}
& 2\gamma  \sum_{z \in \Lambda_L}\rme^{-\ci 2\pi k\cdot z} \pazocal{R}(T_t;x,z) \int_0^t \rmd s\, A_s^{\top}(z) P^{(2)}
 \widehat{A}_s(k) \\\nonumber
& =   2\gamma  \sum_{z \in \Lambda_L}\rme^{-\ci 2\pi k\cdot z} \pazocal{R}(\rme^{-t D}\tau;x,z) \int_0^t \rmd s\, A_s^{\top}(z) P^{(2)}
 \widehat{A}_s(k) \\\nonumber
&\quad + 2\gamma  \sum_{z \in \Lambda_L}\rme^{-\ci 2\pi k\cdot z} \pazocal{R}(\delta_t;x,z) \int_0^t \rmd s\, A_s^{\top}(z) P^{(2)}
 \widehat{A}_s(k)\\\nonumber
 & =:  I^{(5)}_t(x,k) + I^{(6)}_t(x,k).
 \end{align*}
By using \eqref{eq:Ak_estimates}, \eqref{eq:Ax_estimates}, \eqref{eq:delta_tlarge} and \eqref{eq:trivial_R},
\begin{align}
\norm{I_t^{(6)}(x,k)} \lesssim \,  \pazocal{E} Lt^{-3/2} \sum_{z \in \Lambda_L}|z| \rme^{-\gamma_2 |z|}\int_0^t \rmd s\, \rme^{-\delta_0 s} \lesssim \,  \pazocal{E} Lt^{-3/2},
 \end{align} 
 while, thanks to \eqref{eq:Ak_estimates}, \eqref{eq:Ax_estimates}, \eqref{eq:D_est} and \eqref{eq:second_discr_der}, for $ I^{(5)}_t(x,k)$ we have
 \begin{align*}
 \norm{I^{(5)}_t(x,k)} & \lesssim \,  \sum_{z \in \Lambda_L} z^2 \rme^{-\gamma_2 |z|} \int_{\Lambda_L^*} \rmd q \, q^2  \rme^{-t \widehat{D}(q)} |\widehat \tau(q)| \int_0^t \rmd s\, \rme^{-\delta_0 s } \\\nonumber
& \lesssim \,  \pazocal{E} L  \int_{\Lambda_L^*} \rmd q \, q^2  \rme^{-t \widehat{D}(q)} \lesssim \,  \pazocal{E} L\bigg[ \int_{|q| \leq \varepsilon_0} \rmd q \, q^2 \rme^{-t c_1 q^2 }+ \int_{|q| > \varepsilon_0} \rmd q \, q^2 \rme^{-t c_1 \varepsilon_0^2 } \bigg] \\\nonumber
&  \lesssim \,  \pazocal{E} L t^{-3/2}
.
\end{align*}
This guarantees that on the diffusive scale $ I^{(5)} + I^{(6)}=O(L^{-2})$.
Putting together all the terms, we finally get the anticipated equation \eqref{eq:thU_gammalarge}:
\begin{align}\nonumber
U_t(x,k) &=  2 \gamma T_t(x) \int_0^t \rmd s \, \mathcal{A}_s(k) + \ci \mathcal{J}_t(x,k) + O(L^{-2}) \\\nonumber
&=  2 \gamma T_t(x)\int_0^{\infty} \rmd s \, \mathcal{A}_s(k) + \ci 2 \gamma \nabla T_t(x) \int_0^{\infty} \rmd s \, \tilde{\mathcal{A}}_s(k) + O(L^{-2}) \\\nonumber
& = T_t(x) {U}_0(k) +  \ci \nabla T_t(x) {U}_1(k) + O(L^{-2})
\end{align}
where $ \mathcal{J}_t(x,k) =\nabla T_t(x) {U}_1(k) = O(L^{-1})$ on the diffusive scale.
%We note also that one can easily see that $ \mathcal{J}_t(x,k) =O(L^{-1})$ on the diffusive scale.

\section{Kinetic theory of the velocity flip model}\label{sec:Kinetic}

\subsection{Time evolution of the mean Wigner function of normal modes}\label{sec:Wignernew}

In this second part, we are interested in the evolution of a suitably modified Wigner transform $\mathcal{W}_t^{\sigma_1, \sigma_2}(\xi, k)$ 
of phonon normal modes for which we derive a phonon Boltzmann equation.
As in the previous Sections, we deal with dispersion relations which have pinning since then $\omega(k) = \bigl[\widehat{\Phi}(k)\bigr]^{1/2}$ is analytic on a neighbourhood of the real axis and, consequently, its inverse Fourier-transform is an exponentially decreasing function on $\Z$.

It is possible to convert the standard definition of the Wigner function to the lattice setup using distribution techniques to handle points
which lie outside the original lattice \cite{mielke05,ls05}.  We opt here for a different approach: by sacrificing real-valuedness of the Wigner transform, we may continue to consider
it as a function, by using suitable partial Fourier transforms.
More precisely, we consider here
\begin{align}\label{def:mod_Wign}
\mathcal{W}_t^{\sigma_1, \sigma_2}(\xi, k):= \rme^{\ci t \omega(k)(\sigma_1  + \sigma_2)}
\sum_{x \in \Lambda_L} \varphi(\xi-x) \sum_{y \in \Lambda_L} \rme^{-2 \pi \ci y \cdot k}  \E[\psi_t(x,\sigma_1) \psi_t(x+y, \sigma_2)]
\end{align}
where $ \psi_t(x,1)= \psi_t(x)$ and $ \psi_t(x,-1)= \psi^*_t(x)$ are the {\it normal modes} of the harmonic evolution obtained by setting $\gamma$ to zero. 
In Fourier space, they are related to the positions and momenta of the particles by
\begin{align}\label{eq:psi_def}
\widehat \psi_t(k, \sigma)= \frac{1}{\sqrt{2}}(\omega(k) \widehat q (k) + i\sigma \widehat p (k))\, ,
\end{align}
which implies
\begin{align}\label{eq:change_basis}
\widehat q_t(k) = \frac{1}{\sqrt{2} \omega(k)} \sum_{\sigma=\pm 1}\widehat \psi_t(k,\sigma)\, , \qquad 
\widehat p_t(k) = -\frac{\ci}{ \sqrt{2}}\sum_{\sigma=\pm 1} \sigma \widehat \psi_t(k,\sigma)\, .
\end{align}

It is possible to modify the definition of the Wigner function in (\ref{def:mod_Wign})
so that it would enjoy the symmetry properties of the standard Wigner function simply by replacing the factor 
``$\varphi(\xi-x)$'' by ``$\varphi(\xi-x-\frac{1}{2}y)$'': then $(\mathcal{W}_t^{\sigma_1, \sigma_2})^*=\mathcal{W}_t^{-\sigma_2, -\sigma_1}$ and thus $\mathcal{W}_t^{-,+}$ would become real-valued.  However, this choice would make the argument of the testfunction depend on both $x$ and $y$ which would substantially complicate the forthcoming analysis.  Indeed, in what follows we will show %shall be seen soon
that without the $y$-dependence in $\varphi$ the sum over $y$ can be done explicitly, resulting in fairly simple collision operator which is closed under the definition (\ref{def:mod_Wign}).

There does not seem to be any straightforward way of making the Wigner function real without unnecessary
complications.  For instance, $\re \mathcal{W}$ would not satisfy a closed evolution equation.
In addition, the field self-correlation term, $\mathcal{W}_t^{\sigma,\sigma}$, needs the complex factor 
``$ \rme^{\ci t \omega(k) 2 \sigma}$'' to compensate its fast oscillations, resulting in a standard transport term
in the corresponding Boltzmann equation, see (\ref{eq:fullBE}) below.  An additional benefit from the above formulation is that it easily generalizes to higher order cumulants which will become important for evolution problems involving anharmonic potentials \cite[Chapter 4]{Lepri2016}.

In \eqref{def:mod_Wign}, the prefactor $\rme^{\ci t \omega(k)(\sigma_1  + \sigma_2)}$ is needed to cancel out fast oscillations resulting from the free evolution 
for the expectations when a mode is measured against itself, i.e., when $\sigma_1=\sigma_2$.   We also employ a convolution with $\varphi$ to focus on the large scale evolution in space, and we assume that it corresponds to spatial averaging over a given scale $R>0$.  It also provides a map from the discrete values evaluated at $x\in \Lambda_L$
into a smooth function on $\R^d$, $d=1$.
%Here $\varphi$ is a test function which we assume to vary only on the scale $R>0$.
A convenient construction of the test function $\varphi$, which is also well-adapted to the underlying $L$-periodic lattice,
is obtained by taking a Schwartz function $\phi\in \mathcal{S}(\R^d)$, and defining 
\begin{align}\label{eq:test_function}
 \varphi(\xi)= \frac{1}{R^d} \sum_{n \in \Z^d} \phi\!\left(\frac{\xi - L n}{R}\right)\, , \qquad \xi \in \R^d\, .
\end{align}
This definition guarantees that $\varphi$ is smooth, $L$-periodic, and $ \nabla^k_\xi \varphi = O(R^{-k})$ for all $k$.

If it is additionally assumed that $\phi$ is a positive function, normalized to $\int\!\rmd x\, \phi(x)=1$, and that 
its Fourier transform $\widehat{\phi}(k)$ has a compact support, 
we can identify the above convolution with taking of a local average over a region whose 
spatial radius is given by $R$.    
In fact, as shown in Appendix \ref{sec:latticeavk}, as soon as $R$ is greater than the radius of the support of $\FT{\phi}$,
one has $\sum_{x\in \Lambda_L} \varphi(\xi-x) = 1$ for all $\xi\in \R^d$.  Therefore,
for such test functions $\varphi$ the averaging preserves constant densities \defem{exactly}, in the sense 
that it maps constant lattice fields to constant continuum fields without altering the value of the constant.
In the following we shall call test functions $\varphi$ with this property \defem{lattice averaging kernels}.

In this setting the total Hamiltonian energy reads
$$
H = \frac{1}{2} \sum_{\sigma} \int_{\Lambda^*_L}\! \rmd k\,|\widehat \psi(k,\sigma)|^2\,.
$$
Let us point out that the normal mode fields have been normalized so that their 
$\ell_2$-density measures directly the phonon energy; another common choice would be obtained by 
dividing the fields $\widehat \psi$ by $1/\sqrt{\omega(k)}$ in which case the field can be thought of 
as measuring the phonon number density at wavenumber $k$ and each phonon mode carries then an energy $\omega(k)$.

We now define a matrix $M$ as 
\begin{align}\label{def:M_gm_sml}
M(x) = \begin{pmatrix}
0 & \Phi(x) \\
-\cf(x=0) & 0
\end{pmatrix}\,, \quad \text{thus}\quad
\widehat{M}(k) = \begin{pmatrix}
0 & \omega(k)^2 \\
-1 & 0
\end{pmatrix}\, .
\end{align}
We also use the same notation for the translation invariant matrix defined by $M(x,y)=M(x-y)$.
This $M$ is equal to the matrix $M_{\gamma}$, defined earlier in \eqref{def:M_gm_lrg}, evaluated at $\gamma=0$.  

Explicitly, the evolution equation for the position-momentum correlation $C_t$ in (\ref{eq:evolution1}) becomes
\begin{align*}
\partial_t C_t(x,y)=-(M^{\top}C_t)(x,y) - (C_t M)(x,y) - \gamma \left(({J} C_t)(x,y) + (C_t {J})(x,y)\right) + 2 \gamma G_t(x,y)\,,
\end{align*}
where $ {J}(x,y)= \diag (0,\mathbb{1}(x=y))$ and $ G_t (x,y)= \diag (0,\mathbb{1}(x=y)T_t(x) )$, as before. In Fourier space, 
for $\widehat C_t(k_1,k_2)=\sum_{x,y}\rme^{-\ci 2\pi (x k_1+y k_2)} C_t(x,y)$, one has
\begin{align}\label{eq:CtFourier}
 \partial_t \widehat C_t(k_1,k_2)=& %\\\nonumber 
-\widehat M(k_1)^{\top} \widehat C_t(k_1,k_2) - \widehat C_t(k_1,k_2) \widehat M(k_2) \\\nonumber 
&- \gamma (P^{(2)} \widehat C_t(k_1,k_2) + \widehat 
C_t(k_1,k_2) P^{(2)}) + 2 \gamma \widehat T_t(k_1+k_2)P^{(2)} \,.
\end{align}
where $ P^{(2)}= \diag (0,1)$.
Now consider 
$$ 
\E[ \psi_t(x, \sigma_1) \psi_t(y, \sigma_2)] = \int_{(\Lambda_L^*)^2} \rmd k_1 \rmd k_2 \rme^{2 \pi i x \cdot k_1}\rme^{2 \pi i y \cdot k_2} \E [\widehat \psi_t(k_1, \sigma_1) \widehat \psi_t(k_2, \sigma_2)]
$$ where
\begin{align}\label{eq:psi_G}
\E[ \widehat \psi_t(k_1, \sigma_1) \widehat \psi_t(k_2, \sigma_2)] & = \frac{1}{2}[\omega(k_1)\omega(k_2) \widehat C_t^{11}(k_1, k_2) + i \sigma_2 \omega(k_1) \widehat C_t^{12}( k_1, k_2) \\\nonumber 
& \quad + i \sigma_1 \omega(k_2) \widehat C_t^{21}( k_1, k_2)-\sigma_1 \sigma_2 \widehat C_t^{22}( k_1, k_2)] \\\nonumber
& = \Tr [ O(k_1,k_2; \sigma_1, \sigma_2) \widehat C_t(k_1,k_2) ]
\end{align}
with
$$
O(k_1,k_2; \sigma_1, \sigma_2) = \frac{1}{2}
\begin{pmatrix}
\omega(k_1) \omega(k_2) & \ci \sigma_1 \omega(k_2) \\
\ci \sigma_2 \omega(k_1) & - \sigma_1 \sigma_2
\end{pmatrix}\,.
$$
This implies that
\begin{align}\label{eq:def_Y}
Y_t^{\sigma_1, \sigma_2}(x,k)& := \sum_{y \in \Lambda_L } \rme^{-2 \pi \ci y \cdot k} \E [\psi_t(x, \sigma_1) \psi_t(x+y, \sigma_2)] \\\nonumber
& = \int_{\Lambda_L^*} \rmd k' \rme^{2 \pi \ci x \cdot (k+k')} \E [\widehat \psi_t(k', \sigma_1) \widehat \psi_t(k, \sigma_2)]  \\\nonumber
& = \int_{\Lambda_L^*} \rmd k' \rme^{2 \pi \ci x \cdot (k+k')} \Tr [O(k',k; \sigma_1, \sigma_2) \widehat C_t(k',k)]\,,
\end{align}
where $Y_t^{\sigma_1, \sigma_2}(x,k) $ is such that 
\begin{align}\label{eq:WfromY}
\mathcal W_t^{\sigma_1, \sigma_2}(\xi,k) = \rme^{\ci t \omega(k)(\sigma_1+\sigma_2)}\sum_{x \in \Lambda_L} \varphi(\xi-x) Y_t^{\sigma_1, \sigma_2}(x,k).
\end{align}
Then, by using \eqref{eq:CtFourier}, we have
\begin{align}\label{eq:evwigner}
& \partial_t \mathcal{W}_t^{\sigma_1\sigma_2}(\xi,k)= \sum_{x\in\Lambda_L}\varphi (\xi-x)\rme^{\ci t\omega(k) (\sigma_1+\sigma_2)}\int_{\Lambda_L^*} \rmd k' \, \rme^{ 2\pi \ci x\cdot (k+k')} \\\nonumber
& \quad  \times \{\ci (\sigma_1 + \sigma_2) \omega(k) \Tr [O \widehat C_t(k',k)]-\Tr [O \widehat M(k')^{\top} \widehat C_t(k',k) + \widehat M (k) O\widehat C_t(k',k) ]\} \\\nonumber
& - \gamma \sum_{x\in\Lambda_L}\varphi (\xi-x)\rme^{\ci t\omega(k) (\sigma_1+\sigma_2)}\int_{\Lambda_L^*} \rmd k' \, \rme^{ 2\pi \ci x\cdot (k+k')} \\\nonumber
&  \quad \times \Tr [O P^{(2)} \widehat C_t(k',k) + P^{(2)} O  \widehat C_t(k',k)- 2 \widehat T_t(k+k') O P^{(2)} ],
\end{align}
where $ O= O(k',k;\sigma_1,\sigma_2)$. 

We refer to the second term in \eqref{eq:evwigner} as a \defem{collision term}, and denote it by 
\begin{align*}
& \mathcal{C}[\mathcal{W}_t(\xi, \cdot)]^{\sigma_1, \sigma_2}(k) 
= - \gamma \sum_{x\in\Lambda_L}\varphi (\xi-x)\rme^{\ci t\omega(k) (\sigma_1+\sigma_2)}\int_{\Lambda_L^*} \rmd k' \, \rme^{ 2\pi \ci x\cdot (k+k')} \\ 
\nonumber 
& \quad \times \Tr [O P^{(2)} \widehat C_t(k',k) + P^{(2)} O  \widehat C_t(k',k)- 2 \widehat T_t(k+k') O P^{(2)} ] \,,
\end{align*}
where $ O= O(k',k;\sigma_1,\sigma_2)$. 

We first focus on the $\gamma$-independent part. By performing the explicit matrix products we get
\begin{align}\label{eq:flowterm}
& \sum_{x\in\Lambda_L}\varphi (\xi-x)\rme^{\ci t\omega(k) (\sigma_1+\sigma_2)}\int_{\Lambda_L^*} \rmd k' \, \rme^{ 2\pi\ci x\cdot (k+k')} \\\nonumber
& \quad \times \{ \ci (\sigma_1 + \sigma_2) \omega(k) \Tr [O \widehat C_t(k',k)]-\Tr [O \widehat M(k')^{\top} \widehat C_t(k',k) + \widehat M (k) O\widehat C_t(k',k) ]\} \\\nonumber
& = \ci \sigma_1 \sum_{x\in\Lambda_L}\varphi (\xi-x)\rme^{\ci t\omega(k) (\sigma_1+\sigma_2)}\int_{\Lambda_L^*} \rmd k' \, \rme^{ 2\pi \ci x\cdot (k+k')} (\omega(k)-\omega(k'))\Tr [O \widehat C_t(k',k) ]
% \\\nonumber
% & = \ci \sigma_1 \omega(k) \mathcal W_t^{\sigma_1, \sigma_2}(\xi, k)-\ci \sigma_1 \sum_{x\in\Lambda_L}\varphi (\xi-x)\rme^{\ci t\omega(k) (\sigma_1+\sigma_2)}\int_{\Lambda_L^*} \rmd k' \, \rme^{ 2\pi \ci x\cdot (k+k')} \omega(k')\Tr [O \widehat C_t(k',k) ]
\, .
\end{align}
Since $\omega(k')=\omega(-k')$, we may express here
\begin{align*}
 \omega(k)-\omega(k') = \sum_{x'\in \Lambda_L} \tilde{\omega}(x') \rme^{-\ci 2\pi x'\cdot k} \left(1-\rme^{\ci 2\pi x'\cdot (k'+k)}\right)\, ,
\end{align*}
where $\tilde{\omega}$ denotes the inverse discrete Fourier transform of $\omega$ restricted to $\Lambda_L^*$.  Therefore,
\begin{align*}%\label{eq:flowterm}
& \int_{\Lambda_L^*} \rmd k' \, \rme^{ 2\pi \ci x\cdot (k+k')} (\omega(k)-\omega(k'))\Tr [O \widehat C_t(k',k) ] \\ & \quad
= \sum_{x'\in \Lambda_L} \tilde{\omega}(x') \rme^{-\ci 2\pi x'\cdot k} \left(Y_t^{\sigma_1, \sigma_2}(x,k)- Y_t^{\sigma_1, \sigma_2}(x'+x,k)\right)\, .
\end{align*}
Inserting the formula above into 
(\ref{eq:flowterm}) and performing a change a variables in the second term, we obtain that (\ref{eq:flowterm}) is equal to 
\begin{align}%\label{eq:flowterm}
& \ci \sigma_1 \rme^{\ci t\omega(k) (\sigma_1+\sigma_2)} 
\sum_{x,x'\in \Lambda_L} \tilde{\omega}(x') \rme^{-\ci 2\pi x'\cdot k} Y_t^{\sigma_1, \sigma_2}(x,k) 
\left(\varphi(\xi-x)-\varphi(\xi+x'-x)\right)
\, .
\end{align}
Here by Taylor expansion we find that $\varphi(\xi-x)-\varphi(\xi+x'-x) = -x'\cdot\nabla \varphi(\xi -x) + O((x'/R)^2)$.
Then we can replace the discrete sum $\sum_{x'\in \Lambda_L} (-\ci x') \tilde{\omega}(x') \rme^{-\ci 2\pi x'\cdot k}$
with the derivative $\nabla \omega(k)/(2\pi)$ plus a correction which is exponentially small in $L$ due to the exponential 
decay of the Fourier transform of $\omega$ (see the argument in (\ref{eq:discretederivcorrect}) for more details).
Therefore,
\begin{align*}
& \partial_t \mathcal{W}_t^{\sigma_1\sigma_2}(\xi,k)= \sigma_1 \frac{\nabla \omega(k)}{2 \pi}\cdot \nabla_{\xi} \mathcal W_t^{\sigma_1, \sigma_2}(\xi, k)  + 
\mathcal{C}[\mathcal{W}_t(\xi,  \cdot)]^{\sigma_1, \sigma_2}(k)  + O(R^{-2})\, . 
\end{align*}

Let us now come back to the collision term.  Since $O P^{(2)}+P^{(2)} O = O - \diag(\omega(k')\omega(k),\sigma_1\sigma_2)/2$, we have
\begin{align*}
& \mathcal{C}[\mathcal{W}_t(\xi, \cdot)]^{\sigma_1, \sigma_2}(k) \\
& = - \gamma \sum_{x\in\Lambda_L}\varphi (\xi-x)\rme^{\ci t\omega(k) (\sigma_1+\sigma_2)}\int_{\Lambda_L^*} \rmd k' \, \rme^{ 2\pi \ci x\cdot (k+k')} \Tr \bigg[\bigg(O -\frac{1}{2} \diag (\omega(k)\omega(k'),\sigma_1 \sigma_2)\bigg) \widehat C_t(k',k)\bigg] \\\nonumber
& \quad -\sigma_1 \sigma_2 \gamma \sum_{x\in\Lambda_L}\varphi (\xi-x)\rme^{\ci t\omega(k) (\sigma_1+\sigma_2)}\int_{\Lambda_L^*} \rmd k' \, \rme^{ 2\pi \ci x\cdot (k+k')} \widehat T_t(k+k') \\\nonumber
& = - \gamma \mathcal W_t^{\sigma_1, \sigma_2}(\xi, k) \\\nonumber & \quad
+  \frac{\gamma}{2}\sum_{x\in\Lambda_L}\varphi (\xi-x)\rme^{\ci t\omega(k) (\sigma_1+\sigma_2)}\int_{\Lambda_L^*} \rmd k' \, \rme^{ 2\pi \ci x\cdot (k+k')} [\omega(k)\omega(k')\widehat C^{11}_t(k',k) + \sigma_1 \sigma_2 \widehat C^{22}_t(k',k)]  \\\nonumber
& \quad - \sigma_1 \sigma_2 \gamma \sum_{x\in\Lambda_L}\varphi (\xi-x)\rme^{\ci t\omega(k) (\sigma_1+\sigma_2)} \int_{(\Lambda_L^*)^2} \rmd q \rmd q' \,\rme^{2 \pi \ci (q+q') \cdot x} \widehat C^{22}_t(q,q')\,.
\end{align*}
Since
\begin{align*}
\widehat C^{11}_t(k_1,k_2) &= \frac{1}{2 \omega(k_1) \omega(k_2)}\sum_{\sigma_3 \sigma_4} \E [\widehat \psi_t(k_1,\sigma_3)\widehat \psi_t(k_2,\sigma_4)]\, , \\\nonumber
\widehat C^{22}_t(k_1,k_2) &= -\frac{1}{2 }\sum_{\sigma_3 \sigma_4} \sigma_3 \sigma_4\E [\widehat \psi_t(k_1,\sigma_3)\widehat \psi_t(k_2,\sigma_4)]\, ,
\end{align*}
we obtain using \eqref{eq:def_Y} 
\begin{align*}
& \mathcal{C}[\mathcal{W}_t(\xi, \cdot)]^{\sigma_1, \sigma_2}(k)  \\\nonumber
& = - \gamma \mathcal W_t^{\sigma_1, \sigma_2}(\xi, k) +  \frac{\gamma}{4}\sum_{x\in\Lambda_L}\varphi (\xi-x)\rme^{\ci t\omega(k) (\sigma_1+\sigma_2)}\sum_{\sigma_3 \sigma_4}(1-\sigma_1 \sigma_2\sigma_3 \sigma_4) Y^{\sigma_3,\sigma_4}_t(x,k)   \\\nonumber
& \quad + \frac{\sigma_1 \sigma_2 \gamma}{2} \sum_{x\in\Lambda_L}\varphi (\xi-x)\rme^{\ci t\omega(k) (\sigma_1+\sigma_2)} \int_{\Lambda_L^*} \rmd q \sum_{\sigma_3 \sigma_4}\sigma_3 \sigma_4 Y^{\sigma_3,\sigma_4}_t(x,q) \\\nonumber
& = - \gamma \mathcal W_t^{\sigma_1, \sigma_2}(\xi, k) +  \frac{\gamma}{4}\sum_{\sigma_3 \sigma_4}(1-\sigma_1 \sigma_2\sigma_3 \sigma_4) \rme^{\ci t\omega(k) (\sigma_1+\sigma_2-\sigma_3-\sigma_4)} \mathcal W^{\sigma_3,\sigma_4}_t(\xi,k)   \\\nonumber
& \quad + \frac{\gamma }{2} \sum_{\sigma_3 \sigma_4}\sigma_1 \sigma_2\sigma_3 \sigma_4 \int_{\Lambda_L^*} \rmd q \, \rme^{\ci t[\omega(k) (\sigma_1+\sigma_2)-\omega(q) (\sigma_3+\sigma_4)]} \mathcal W^{\sigma_3,\sigma_4}_t(\xi,q) \, .
\end{align*}
Expanding the various sign combinations explicitly thus yields 
%Therefore we can rewrite the collision operator \eqref{coll_op} in a more explicit form
\begin{align}\label{eq:Coll1}
& \mathcal{C}[\mathcal{W}_t(\xi, \cdot)]^{\sigma_1, \sigma_2}(k) 
 = - \gamma \mathcal W_t^{\sigma_1, \sigma_2}(\xi, k) \\\nonumber 
& \quad +  \frac{\gamma}{4}(1-\sigma_1 \sigma_2) \rme^{\ci t\omega(k) (\sigma_1+\sigma_2)} (\rme^{-2\ci t\omega(k)} \mathcal W^{+,+}_t(\xi,k) + \rme^{2\ci t\omega(k)}\mathcal W^{-,-}_t(\xi,k) ) \\\nonumber 
& \quad + \frac{\gamma}{4}(1 + \sigma_1 \sigma_2) \rme^{\ci t\omega(k) (\sigma_1+\sigma_2)}(\mathcal W^{+,-}_t(\xi,k) + \mathcal W^{-,+}_t(\xi,k)) \\\nonumber
& \quad + \frac{\gamma }{2} \sigma_1 \sigma_2 \rme^{\ci t \omega(k) (\sigma_1+\sigma_2)} \int_{\Lambda_L^*} \rmd q (\rme^{-2 \ci t \omega(q) } \mathcal W^{+,+}_t(\xi,q) + \rme^{2 \ci t \omega(q) } \mathcal W^{-,-}_t(\xi,q)) \\\nonumber
& \quad - \frac{\gamma }{2} \sigma_1 \sigma_2 \rme^{\ci t\omega(k) (\sigma_1+\sigma_2)} \int_{\Lambda_L^*} \rmd q ( \mathcal W^{+,-}_t(\xi,q) + \mathcal W^{-,+}_t(\xi,q))\,.
\end{align}

If $\varphi$ is real-valued, as we assume here, the components of $\mathcal{W}$ can be related to each other by complex conjugation.  Namely, then
\begin{align}\label{W_tr12}
 \mathcal{W}_t^{\sigma_1, \sigma_2}(\xi, k) &
% = \sum_{x \in \Lambda_L} \varphi(\xi-x) \sum_{y \in \Lambda_L} \rme^{-2 \pi \ci y \cdot k} \rme^{\ci t \omega(k)(\sigma_1  + \sigma_2)} \E[\psi_t(x,\sigma_1) \psi_t(x+y, \sigma_2)] \\\nonumber 
% & 
= \bigg( \sum_{x,y \in \Lambda_L}\varphi(\xi-x)  \rme^{2 \pi \ci y \cdot k} \rme^{-\ci t \omega(k)(\sigma_1  + \sigma_2)} \E[\psi_t(x, -\sigma_1)\psi_t(x+y,-\sigma_2) ] \bigg)^* \\\nonumber
& = (\mathcal{W}_t^{-\sigma_1, -\sigma_2}(\xi, -k))^* \, .
\end{align}
In addition, from the regularity properties of the test function we can also estimate the effect of swapping the sign of $k$ and the order of the $\sigma$-indices:
making a change of variables from $x$ to $x'=x+y$ it follows that
\begin{align}\label{W_tr1}
& \mathcal{W}_t^{\sigma_1, \sigma_2}(\xi, k)= 
% \sum_{x \in \Lambda_L} \varphi(\xi-x) \sum_{y \in \Lambda_L} \rme^{-2 \pi \ci y \cdot k} \rme^{\ci t \omega(k)(\sigma_1  + \sigma_2)} \E[\psi_t(x,\sigma_1) \psi_t(x+y, \sigma_2)] \\\nonumber 
% & =
\sum_{x',y \in \Lambda_L} \varphi(\xi-x' + y)  \rme^{-2 \pi \ci y \cdot k} \rme^{\ci t \omega(k)(\sigma_1  + \sigma_2)} \E[\psi_t(x', \sigma_2)\psi_t(x'-y,\sigma_1) ] \\\nonumber
% & = \sum_{x',y \in \Lambda_L}  \varphi(\xi-x' - y)  \rme^{2 \pi \ci y \cdot k} \rme^{\ci t \omega(k)(\sigma_1  + \sigma_2)} \E[\psi_t(x', \sigma_2)\psi_t(x'+y,\sigma_1) ] \\\nonumber
& = \sum_{x',y \in \Lambda_L} (\varphi(\xi-x')+\varphi(\xi-x' - y)-\varphi(\xi-x'))  \rme^{2 \pi \ci y \cdot k} \rme^{\ci t \omega(k)(\sigma_1  + \sigma_2)} \E[\psi_t(x', \sigma_2)\psi_t(x'+y,\sigma_1) ] \\\nonumber
& = \mathcal{W}_t^{\sigma_2, \sigma_1}(\xi, -k) + O(R^{-1})\,.
\end{align}
This second formula, however, needs to be used with some care since the correction might not be bounded in the lattice size $L$.
This is guaranteed if the correlations decay fast enough in space so that $\sum_y |y| |\E[\psi_t(x', \sigma_2)\psi_t(x'+y,\sigma_1) ]|$ remains bounded in $L$.
Whenever this is the case, we can combine the above bounds and conclude that
\begin{align}\label{W_tr2}
\mathcal{W}_t^{\sigma_1, \sigma_2}(\xi, k) = (\mathcal{W}_t^{-\sigma_2, -\sigma_1}(\xi,k))^* + O(R^{-1}).
\end{align}

The closest quantity to the standard Wigner function is the function $\mathcal{W}_t^{-,+}(\xi, k)$.
By \eqref{W_tr2}, it satisfies $\mathcal{W}_t^{-,+}(\xi, k) = (\mathcal{W}_t^{-,+}(\xi,k))^* + O(R^{-1})$ whenever the correlations decay sufficiently rapidly.
Therefore, although this function is not necessarily real, its imaginary part is typically very small, due to the spatial averaging.

By using \eqref{W_tr12} and the symmetry of $\omega$ we find from \eqref{eq:Coll1}
\begin{align}%\label{eq:Coll2}
& \mathcal{C}[\mathcal{W}_t(\xi, \cdot)]^{\sigma_1, \sigma_2}(k) 
 = - \gamma \mathcal W_t^{\sigma_1, \sigma_2}(\xi, k) \\\nonumber 
& \quad +  \frac{\gamma}{4}(1-\sigma_1 \sigma_2) \rme^{\ci t\omega(k) (\sigma_1+\sigma_2)} (\rme^{-2\ci t\omega(k)} \mathcal W^{-,-}_t(\xi,-k)^* + \rme^{2\ci t\omega(k)}\mathcal W^{-,-}_t(\xi,k) ) \\\nonumber 
& \quad + \frac{\gamma}{4}(1 + \sigma_1 \sigma_2) \rme^{\ci t\omega(k) (\sigma_1+\sigma_2)}(\mathcal W^{-,+}_t(\xi,-k)^* + \mathcal W^{-,+}_t(\xi,k)) \\\nonumber
& \quad + \gamma \sigma_1 \sigma_2 \rme^{\ci t \omega(k) (\sigma_1+\sigma_2)} \int_{\Lambda_L^*}\! \rmd q \,
 \re\!\left[ \rme^{2 \ci t \omega(q) } \mathcal W^{-,-}_t(\xi,q) - \mathcal W^{-,+}_t(\xi,q)\right]\,.
\end{align}
Then the equal sign term is given by
\begin{align}\label{eq:Collmm}
& \mathcal{C}[\mathcal{W}_t(\xi, \cdot)]^{-,-}(k) 
 = - \gamma \mathcal W_t^{-,-}(\xi, k) 
% \\\nonumber & \quad 
 + \frac{\gamma}{2} \rme^{-\ci 2 t \omega(k)}(\mathcal W^{-,+}_t(\xi,-k)^* + \mathcal W^{-,+}_t(\xi,k)) \\\nonumber
& \quad + \gamma \rme^{-\ci 2 t \omega(k)} \int_{\Lambda_L^*}\! \rmd q \,
 \re\!\left[ \rme^{2 \ci t \omega(q) } \mathcal W^{-,-}_t(\xi,q) - \mathcal W^{-,+}_t(\xi,q)\right]\,,
\end{align}
and the opposite sign term by 
\begin{align}\label{eq:Coll2}
& \mathcal{C}[\mathcal{W}_t(\xi, \cdot)]^{-,+}(k) 
 = - \gamma \mathcal W_t^{-,+}(\xi, k) 
 +  \frac{\gamma}{2} (\rme^{-2\ci t\omega(k)} \mathcal W^{-,-}_t(\xi,-k)^* + \rme^{2\ci t\omega(k)}\mathcal W^{-,-}_t(\xi,k) ) \\\nonumber 
& \quad + \gamma  \int_{\Lambda_L^*}\! \rmd q \,
 \re\!\left[\mathcal W^{-,+}_t(\xi,q) -\rme^{2 \ci t \omega(q) } \mathcal W^{-,-}_t(\xi,q)\right]\,.
\end{align}
Hence, these two functions satisfy a closed pair of evolution equations of the form
\begin{align}\label{eq:fullBE}
& \partial_t \mathcal{W}_t^{-,+}(\xi,k) + v(k)\cdot \nabla_{\!\xi} \mathcal W_t^{-,+}(\xi, k) + O(R^{-2}) =
\mathcal{C}[\mathcal{W}_t(\xi, \cdot)]^{-,+}(k) \, ,\\ \nonumber 
& \partial_t \mathcal{W}_t^{-,-}(\xi,k) + v(k)\cdot \nabla_{\!\xi} \mathcal W_t^{-,-}(\xi, k) + O(R^{-2}) =
\mathcal{C}[\mathcal{W}_t(\xi, \cdot)]^{-,-}(k) \, , 
\end{align}
where both ``bands'' have a phonon velocity $v(k)=\frac{\nabla \omega(k)}{2 \pi}$.

Let us stress that no approximations have been made to get to the above pair of equations, and they are valid for all scale parameters $R>0$, as long as the testfunction $\varphi$ is constructed as mentioned in the beginning of this section.  Of course, to be of any use as a transport equation, one needs to make sure that the effect of the correction terms marked as ``$O(R^{-2})$'' above remains small.  What is commonly done in mathematical derivations of kinetic equations is to scale also time by $R$ and then take $R\to \infty$ in such a way that the collision operator on the right hand side has a finite nontrivial limit.  In the present case, this could be achieved by taking $t=\tau R$ and $R=\gamma^{-1}$ and then considering a weak noise limit  $\gamma\to 0$ for a fixed $\tau>0$. This would correspond to the standard kinetic scaling limit, and we refer to \cite{Basile2009,Jara2015} for methods of controlling the limit rigorously in similar stochastic systems.

The kinetic scaling limit however hides a difficulty whose solution begs for an explanation: the above computation shows that the correction 
term $O(R^{-2})$ would be present even for pure  \defem{harmonic} evolution.  It is in fact a necessary term which captures the difference between transport by the discrete wave equation and its radiative transport approximation obtained by setting the $O(R^{-2})$ term to zero.   Apparently, then the kinetic equation is accurate only up to times $t=O(R^2)$ which with above kinetic scales would mean $t=O(\gamma^{-2})$.  
However, kinetic theory does correctly predict the leading contribution to thermal conductivity in a number of phonon systems, and in the rest of this section we will show that this is also the case for the present velocity flip model.  In fact, in this special case, the kinetic prediction turns out to be exact, and as proven in \cite{L14,Simon12}, diffusion of energy persists for all sufficiently large times and describes correctly the $t\to\infty$ asymptotics of the energy density.

To reconcile the apparent restriction of kinetic theory to times $t\lesssim R^{2}$
with the fact that it does correctly capture even $t\to\infty$ asymptotics, we deviate here from the standard kinetic limit approach to kinetic equations by introducing a new spatial scale $R$ to its definition.  We do not specify the value of $R$ exactly in the following, merely assume that it is sufficiently large that certain homogenization properties hold.  In particular, we will assume that $R$ is much larger than the mean free path of phonons but not so large that it washes out macroscopic effects: we assume that  $\gamma^{-1}\ll R \lesssim L$.

\subsection{Stationary solutions}\label{sec:stationary}

The left-hand sides in \eqref{eq:fullBE} include a time-derivative and a transport  term of order $O(R^{-1})$ while the collision 
terms on the right hand side are $O(\gamma)$ for small $\gamma$.  Qualitatively, the equation corresponds to phonons with wavenumber $k$ moving at a velocity $v(k)$ and experiencing collisions at a rate $O(\gamma)$.  Thus the mean free path of phonons should have a magnitude
$ \gamma^{-1}|v(k)|$.  Therefore, for spatial scales much larger than the mean free path, i.e., whenever 
$R\gg \gamma^{-1}|v(k)|$,
the time evolution of the above Wigner functions is dominated by the right hand side, i.e.\ the collision term. 
In that case, it is reasonable to start by first solving the equation 
including only the effect of collisions.  Since the collisions do not mix values with different $\xi$, this amounts to solving the equations 
\eqref{eq:fullBE} for spatially homogeneous initial data.  For this reason, let us suppose in this subsection that the point $\xi$ is a fixed parameter which we drop from the notation.

We now want to find the stationary solution of the above system 
in the translation invariant case.  We define the following quantities:
\begin{align}\label{eq:defHIPQ}
\pazocal{H}_t(k) & := \frac{1}{2}(\mathcal{W}_t^{-,+}(k) + \mathcal{W}_t^{-,+}(-k)^*)\, , \\\nonumber
\pazocal{I}_t(k) & := \frac{1}{2}(\mathcal{W}_t^{-,+}(k) - \mathcal{W}_t^{-,+}(-k)^*)\, , \\\nonumber
\pazocal{P}_t(k) & := \frac{1}{2}(\rme^{2 \ci t \omega(k)}\mathcal{W}_t^{-,-}(k) + \rme^{-2 \ci t \omega(k)}\mathcal{W}_t^{-,-}(-k)^*)\, , \\\nonumber
\pazocal{Q}_t(k) & := \frac{1}{2}(\rme^{2 \ci t \omega(k)}\mathcal{W}_t^{-,-}(k) - \rme^{-2 \ci t \omega(k)}\mathcal{W}_t^{-,-}(-k)^*)\, . 
\end{align}
Each of these functions is either symmetric ($\pazocal{H}, \pazocal{P}$) or antisymmetric ($\pazocal{I}, \pazocal{Q}$) under the transform
$F(k) \to F(-k)^*$.  They allow writing the collision operator in a very compact form.  Namely, 
by introducing the simplified collision operator $ \bar{\mathcal C}$, defined as
\begin{align}\label{eq:simpleColl}
\bar{\mathcal C} [f](k):= \gamma \int_{\Lambda_L^*} \rmd q \, [f(q)-f(k)]\, ,
\end{align}
we find from \eqref{eq:fullBE} the homogeneous evolution equations
\begin{align}\label{eq:H}
\partial_t \pazocal H_t(k) & = \bar{\mathcal C}[\pazocal{H}_t-\pazocal{P}_t](k)\, , \\\nonumber
\partial_t \pazocal I_t (k) & = -\gamma \pazocal{I}_t(k)\, .
\end{align}
In particular, then $ \pazocal I_t (k) = \pazocal I_0(k) \rme^{-\gamma t}$ and it approaches the unique stationary solution
$\pazocal I (k)=0$ exponentially fast as soon as $t=O(\gamma^{-1})$.  

The homogeneous equations for $\pazocal{P}, \pazocal{Q}$ are slightly more complicated, namely
\begin{align}
\partial_t 
\begin{pmatrix}
\pazocal P_t(k) \\
\pazocal Q_t(k)
\end{pmatrix}
= \pazocal{L}
\begin{pmatrix}
\pazocal P_t(k) \\
\pazocal Q_t(k)
\end{pmatrix}
+ \begin{pmatrix}
\gamma\int_{\Lambda_L^*} \rmd q \, \pazocal P_t(q) \\
0
\end{pmatrix}
-  \begin{pmatrix}
\bar{\mathcal C}[\pazocal{H}_t](k) \\
0
\end{pmatrix}
\end{align}
where
$$
\pazocal{L} := \begin{pmatrix}
-\gamma & 2 \ci \omega(k) \\
2 \ci \omega(k) & - \gamma
\end{pmatrix}
$$
and its eigenvalues are $ \lambda = -\gamma \pm 2 \ci \omega(k)$. The stationary solutions then satisfy
\begin{align}
 \begin{pmatrix}
\pazocal P(k) \\
\pazocal Q(k)
\end{pmatrix}
= - \pazocal{L}^{-1}\begin{pmatrix}
\gamma \int_{\Lambda_L^*} \rmd q \, \pazocal P(q) -\bar{\mathcal C}[\pazocal{H}](k)\\
0
\end{pmatrix}
.
\end{align}
Since 
$$
\pazocal{L}^{-1} = - \frac{1}{\gamma^2 + 4 \omega^2(k)} \begin{pmatrix}
\gamma & 2 \ci \omega(k) \\
2 \ci \omega(k) & \gamma
\end{pmatrix},
$$
one gets
\begin{align}\label{eq:PQ}
 \begin{pmatrix}
\pazocal P(k) \\
\pazocal Q(k)
\end{pmatrix}
= \frac{1}{\gamma^2 + 4 \omega^2(k)} \begin{pmatrix}
\gamma & 2 \ci \omega(k) \\
2 \ci \omega(k) & \gamma
\end{pmatrix}\begin{pmatrix}
\gamma \int_{\Lambda_L^*} \rmd q \, \pazocal P(q) - \bar{\mathcal C}[\pazocal{H}](k) \\
0
\end{pmatrix}
.
\end{align}
We observe that the stationary equation corresponding to \eqref{eq:H} is $\bar{\mathcal{C}}[\pazocal H - \pazocal P] =0$. 
Then, since $ \bar{\mathcal{C}}$ is linear, we have 
\begin{align}\label{eq:collPH}
\bar{\mathcal{C}}[\pazocal H]= \bar{\mathcal{C}}[ \pazocal P] \,.
\end{align}
Thus from \eqref{eq:simpleColl}, \eqref{eq:PQ} and \eqref{eq:collPH} the equation for $ \pazocal P(k)$ becomes
\begin{align}
\pazocal P(k) = \frac{\gamma}{\gamma^2 + 4 \omega^2(k)} \bigg[\gamma \int_{\Lambda_L^*} \rmd q \, \pazocal P(q)- \bar{\mathcal C}[\pazocal{P}](k)  \bigg] = \frac{\gamma^2}{\gamma^2 + 4 \omega^2(k)}\pazocal P(k).
\end{align}
Since $\gamma^{2}/(\gamma^2 + 4 \omega^2(k))<1$ for $\omega(k)>0$, then necessarily $ \pazocal P(k) =0$.
Therefore, by \eqref{eq:collPH} we now get $ \bar{\mathcal{C}}[\pazocal H] =0$ and consequently also $ \pazocal Q(k) =0$. 

The equation $ \bar{\mathcal{C}}[\pazocal H] =0$ is solved precisely by functions which are constant in $k$.
Therefore, we have now proven that to each stationary solution there is a constant $E$ 
such that $ \pazocal{H}(k)=E$ and 
$ 0=\pazocal P(k) = \pazocal Q(k) = \pazocal I(k)$.  In addition, then clearly
\begin{align}\nonumber
\mathcal{W}^{-,-}(k) & = 0\, , \\\nonumber
\mathcal{W}^{-,+}(k) & = \pazocal{H}(k)  = \mathcal{W}^{-,+}(-k)^*\, ,
\end{align}
where $ \mathcal{W}^{-,-}(k)$ and $ \mathcal{W}^{-,+}(k)$ denote the stationary counterparts of $ \mathcal{W}_t^{-,-}(k)$  and $ \mathcal{W}_t^{-,+}(k)$.
The second equality implies also that 
$\pazocal{H}(k) = E$ is a \defem{real} constant.

\subsection{Boltzmann equation for the energy density}

As we already observed before, 
the definition of $ \mathcal W_t(\xi,k)$ indicates 
that $ \mathcal{W}_t^{-,+}(\xi,k)$ is the quantity closest 
to the standard Wigner transform.   Thus we would expect it to be of special interest
in the kinetic theory; let us denote 
\begin{align}
 W_t(\xi,k) & := \mathcal{W}_t^{-,+}(\xi,k)\, .
\end{align} 
The relaxation of $W$ is then governed by the phonon Boltzmann equation 
\begin{align}\label{eq:WpBE}
  \partial_t W_t(\xi,k) + v(k)\cdot \nabla_{\xi} W_t(\xi,k) &= \bar{\mathcal{C}}[W_t(\xi,\cdot)](k)\, ,
\end{align}
where we have
used the simplified collision operator $\bar{\mathcal{C}}$ defined in (\ref{eq:simpleColl}).  This equation follows from (\ref{eq:fullBE}) after we 
assume that equilibration %thermalization 
is so fast that 
both $\bar{\mathcal{C}}[\pazocal{P}_t]$ and the difference between $W_t$ and $\pazocal{H}_t$ can be neglected (note that 
$W_t-\pazocal{H}_t=\pazocal{I}_t$, and thus it relaxes to zero independently from the other fields).

Moreover, assuming that $R\gg \gamma^{-1}$, 
for times $t=O(R)$, i.e.\ after the collisions have had plenty of time to push the system towards equilibrium, 
we expect that to every $\xi$ there should be a real constant 
$E_t(\xi)$ such that $W_t(\xi,k)-E_t(\xi)$ is small.  By construction, 
$W_t$ is a function which varies only at the scale $R$ in $\xi$, i.e.\ $\nabla_\xi W_t = O(R^{-1})$,
and thus $E_t(\xi)$ should then also be similarly slowly varying.

In order to find $ E_t(\xi)$, we integrate the definition (\ref{def:mod_Wign}), so that 
$$
\int_{\Lambda_L^*} \rmd k \, W_t(\xi,k) 
= \sum_{x \in \Lambda_L} \varphi(\xi-x) \E[|\psi_t(x)|^2]
= \int_{\Lambda_L^*} \rmd k \, \pazocal H_t(\xi,k)
$$ 
which is clearly nonnegative 
for nonnegative  testfunctions $\varphi$.  
If $\varphi$ is one of the ``lattice averaging kernels'' discussed in Section \ref{sec:Wignernew}, we also have
$$\int_{-L/2}^{L/2}\rmd \xi \sum_{x \in \Lambda_L} \varphi(\xi-x) \E[|\psi_t(x)|^2] = H_L(q_t,p_t)=H_L(q_0,p_0),$$ and thus 
then we may identify the constant
$E_t(\xi):=\int_{\Lambda_L^*} \rmd k \, W_t(\xi,k)$ as the energy in a volume of radius $R$ centered at point $\xi$, i.e.\
it is equal to the physical energy density at $\xi$ at the time $t$.  Furthermore, this implies that $ W_t(\xi,k)$ can be 
interpreted as the ``density''\footnote{This function is not necessarily positive, hence the quotation marks here.} in the phonon phase space with variables $(\xi,k) $, similarly to how 
the standard Wigner transform achieves the goal in quantum mechanics. Let us stress that, since the definition involves taking expectation 
over the randomness, this refers to the energy density averaged over realizations of the velocity flips.

\subsection{Kinetic theory prediction for diffusion of  energy}\label{sec:conductivity}

The Boltzmann equation (\ref{eq:WpBE}) also allows studying the relaxation towards global equilibrium.  This is one of the standard uses
of kinetic theory, and we merely recall here how the argument works in the present case, giving only heuristic justification for the various steps.  As mentioned in the Introduction, diffusion of
energy in the present velocity flip model at standard hydrodynamics scales has already been rigorously proven in \cite{L14,Simon12}.  We hence
skip any rigorous estimates, and focus on trying to understand how the known diffusion phenomena is connected to the above kinetic equation.  As a byproduct, we obtain a simple integral formula for the thermal conductivity which is shown to coincide with the previous results, at least in the special case of nearest neighbour interactions for which the integral can be computed analytically in the limit $L\to \infty$.

Let us suppose that the final phase of equilibration occurs via processes which are slower than ballistic, in which case
$\partial_t W_t$ is small compared to $v(k)\cdot \nabla_{\xi} W_t$.  This would occur for instance if the relaxation is diffusive,
since then densities averaged over a volume of radius $O(R)$ change at a rate $O(R^{-2})$, and thus then
$\partial_t W_t=O(R^{-2})$ and $v(k)\cdot \nabla_{\xi} W_t=O(R^{-1})$.

Therefore, combined with the earlier relaxation argument, for such systems we expect that 
$$ {W}_t(\xi,k) =  E_t(\xi) + \epsilon_t(\xi,k) \, , $$
where $\epsilon_t$ is small and by the definition of $E_t(\xi)$ we have 
%$\epsilon_t(\xi,k)=W_t(\xi,k)-E_t(\xi)=\gamma^{-1} \bar{\mathcal{C}}[W_t(\xi,\cdot)](k)$. In particular, 
$\int_{\Lambda_L^*} \rmd k \,\epsilon_t(\xi,k)=0$.
Since $\partial_t W_t$ is assumed to be of lower order, the dominant part of $\epsilon_t$ can be found by solving
the equation
\begin{align}\label{eq:stat1}
v(k)\nabla_{\xi}{W}_t(\xi,k) \simeq \bar{\mathcal{C}}[{W}_t(\xi,\cdot )](k)\, .
\end{align}

In the general version of the argument, which can be found for instance in Sec.\ 14 of \cite{spohn05}, one then proceeds by using the expansion
$\bar{\mathcal{C}}[{W}_t(\xi,\cdot )](k) = \mathcal{L}_{E_t(\xi)}[\epsilon_t(\xi,\cdot )](k) + O(\epsilon_t^2)$
where $\mathcal{L}_{E}$ denotes the linearization of the collision operator $\bar{\mathcal{C}}$ around the stationary solution $E$.
Then the dominant perturbation can be found by applying the inverse $\mathcal{L}_{E_t(\xi)}^{-1}$ to (\ref{eq:stat1}).

In the present case, the collision operator is not only linear---which always implies that the linearized operator is the same as the original collision operator---but it is in fact a very simple projection operator.  The inverse is explicit and for our purposes can be found directly from the definition of $\epsilon_t$.  Namely, since $E_t(\xi)= \int_{\Lambda_L^*} \rmd k \,W_t(\xi,k)$, 
we have $$\bar{\mathcal{C}}[{W}_t(\xi,\cdot )](k) = \gamma (E_t(\xi)-{W}_t(\xi,k)) = -\gamma \epsilon_t(\xi,k).$$
On the other hand, the dominant part of $v(k)\nabla_{\xi}{W}_t(\xi,k)$ is given by $v(k)\nabla_{\xi}{E}_t(\xi)$, and thus 
\eqref{eq:stat1} implies that 
\begin{align}\label{eq:epsilon}
\epsilon(\xi,k)\simeq -\gamma^{-1} v(k) \nabla_{\xi} E_t(\xi) .  
\end{align}

This result can be connected with the energy flux by using the conservation law which is reflected in the identity
$\int_{\Lambda_L^*} \rmd k \, \bar{\mathcal{C}}[W](k) = 0$, valid for any function $W$.
Thus for any solution of \eqref{eq:WpBE} we have 
\begin{align*}
\partial_t E_t(\xi) =  \int_{\Lambda_L^*} \rmd k \left(-v(k)\nabla_{\xi}{W}_t(\xi,k) \right) = -\nabla_{\xi} j_t(\xi)\, ,
\end{align*}
where
\begin{align*}
j_t(\xi) := &  \int_{\Lambda_L^*}\! \rmd k \, v(k) {W}_t(\xi,k) 
\end{align*}
can be identified as the energy current.  At equilibrium, for $W_t(\xi,k)=E_t(\xi)$, the flux vanishes, since 
$v(-k)=-v(k)$ due to the symmetry of the dispersion relation $\omega$.  Therefore, we can now conclude that the 
energy current satisfies 
$j_t(\xi) =   \int_{\Lambda_L^*}\! \rmd k \, v(k) \epsilon_t(\xi,k)$.
Together with \eqref{eq:epsilon} this implies that, under the above assumptions about the relaxation process,
the dominant part of the energy flux is given by 
\begin{align*}
j_t(\xi) \simeq -\kappa \nabla_{\xi} E_t(\xi)\, ,
\end{align*}
where
\begin{align}\label{eq:defkappal}
\kappa = \kappa(L) := \gamma^{-1}\int_{\Lambda_L^*} \rmd k \, v(k)^2  \, .
\end{align}
Inserting the approximation into the continuity equation then results in the equation 
\begin{align*}
\partial_t E_t(\xi)  \simeq  \kappa \nabla_{\xi}^2{E}_t(\xi) \, ,
\end{align*}
which is a linear diffusion equation with a diffusion constant $\kappa$.  This in fact implies that if the assumption about eventual
slow relaxation holds, then energy density must relax diffusively, with a diffusion constant $\kappa$.  

Finally, let us point out that this formula coincides with the conductivity obtained from the nonequilibrium steady state current 
of the system with the same bulk dynamics but with heat baths at the two ends enforcing a steady state current through the system.
As in the references, suppose that the harmonic interactions connect only the nearest neighbours and have the dispersion relation 
$\omega(k)=\sqrt{\omega_0^2+4 \sin^2(\pi k)}$, with $\omega_0>0$. 
As shown in \cite{DKL11}, the steady state covariance matrix is then identical to the one of 
the so called self-consistent heat bath model.  The self-consistent model was studied in detail in \cite{bll02}, and 
its thermal conductivity is given in Equation (4.18) of the reference.  
As shown a few lines above the formula, in Equation (4.16), the conductivity may be represented by a one-dimensional integral as
\begin{align}
 \kappa[\text{Ref.~\cite{bll02}}] = \frac{1}{\gamma}\int_0^1 \!\rmd x\, \frac{\sin^2(\pi x)}{\omega_0^2+4 \sin^2(\pi x/2)}\, .
\end{align}
Since in this case $v(k)=\omega'(k)/(2\pi)=\sin(2\pi k)/\omega(k)$, after employing evenness of the integrand and performing
a change of variables to $x=2 k$, the result clearly
coincides with the $L\to\infty$ limit of $\kappa(L)$ given in Eq.\ (\ref{eq:defkappal}) above.

\subsection{Kinetic theory prediction for particle correlations}\label{sec:kin_corr}

To inspect the accuracy of the above discussion in more detail, let us derive a prediction about the structure of  the 
$q_x(t), p_x(t)$ covariance matrix and compare this to the earlier results derived using the exact solution of its evolution.
To facilitate the comparison, let us next consider 
the Wigner function of the position-momentum correlation matrix $\mathcal{U}_t(\xi,k)$ which we define analogously to 
$ \mathcal{W}_t(\xi,k)$ using the formula
\begin{align}\label{eq:U_new}
\mathcal{U}_t(\xi,k)= \sum_{x,y \in \Lambda_L} \varphi(\xi -x)\rme^{-2 \pi \ci y k} C_t(x,x+y)\, .
\end{align}
It is a spatially averaged version of the matrix function $U_t(x,k)$ defined earlier in \eqref{eq:U_wigner}.

The change of basis formula \eqref{eq:change_basis} then yields 
\begin{align} \nonumber
\mathcal{U}^{11}_t(\xi,k) & =  \frac{1}{2} \sum_{x \in \Lambda_L} \varphi(\xi-x)\int_{\Lambda_L^*} \rmd k' \, \frac{\rme^{2 \pi \ci x (k+k')}}{\omega(k)\omega(k')} \sum_{\sigma,\sigma'}\E[\widehat \psi_t(k',\sigma') \widehat \psi_t(k,\sigma)]\, , \\\nonumber
 \mathcal{U}^{12}_t(\xi,k) & =  - \frac{\ci}{2} \sum_{x \in \Lambda_L} \varphi(\xi-x) \int_{\Lambda_L^*} \rmd k' \, \frac{\rme^{2 \pi \ci x (k+k')}}{\omega(k')} \sum_{\sigma,\sigma'}\sigma \E[\widehat \psi_t(k',\sigma') \widehat \psi_t(k,\sigma)]\, ,\\\nonumber
 \mathcal{U}^{21}_t(\xi,k) & =  -\frac{\ci}{2} \sum_{x \in \Lambda_L} \varphi(\xi-x) \int_{\Lambda_L^*} \rmd k' \, \frac{\rme^{2 \pi \ci x (k+k')}}{\omega(k)} \sum_{\sigma,\sigma'}\sigma'\E[\widehat \psi_t(k',\sigma') \widehat \psi_t(k,\sigma)]\, , \\\nonumber
 \mathcal{U}^{22}_t(\xi,k) & =  \frac{1}{2} \sum_{x \in \Lambda_L} \varphi(\xi-x) \int_{\Lambda_L^*} \rmd k' \, \rme^{2 \pi \ci x (k+k')} 
\sum_{\sigma,\sigma'}(-\sigma'\sigma) \E[\widehat \psi_t(k',\sigma') \widehat \psi_t(k,\sigma)] \, .
 \end{align} 
For $\mathcal{U}^{21}$ and $\mathcal{U}^{22}$ we obtain immediately from \eqref{eq:def_Y}, \eqref{eq:WfromY}, and \eqref{W_tr12}
\begin{align}
\mathcal{U}^{21}_t(\xi,k)& =\frac{\ci}{2 \omega(k)}\left(\mathcal{W}_t^{-,+}(\xi,k)-\mathcal{W}_t^{+,-}(\xi,k)
+\rme^{2 \ci t \omega(k)} \mathcal{W}_t^{-,-}(\xi,k)-\rme^{-2 \ci t \omega(k)} \mathcal{W}_t^{+,+}(\xi,k)\right) \nonumber\\
% & = \frac{\ci}{2 \omega(k)}(\mathcal{W}_t^{-,+}(\xi,k)-\mathcal{W}_t^{+,-}(\xi,k) + \mathcal{W}_t^{-,+}(\xi,-k)- \mathcal{W}_t^{-,+}(\xi,-k)) \\\nonumber
& = \frac{\ci}{\omega(k)}(\pazocal{I}_t(\xi,k)-\pazocal{Q}_t(\xi,k))\, ,
\\\nonumber
\mathcal{U}^{22}_t(\xi,k)& = \frac{1}{2}\left(\mathcal{W}_t^{-,+}(\xi,k)+\mathcal{W}_t^{+,-}(\xi,k)
-\rme^{2 \ci t \omega(k)} \mathcal{W}_t^{-,-}(\xi,k)-\rme^{-2 \ci t \omega(k)} \mathcal{W}_t^{+,+}(\xi,k)\right) \nonumber\\
& = \pazocal{H}_t(\xi,k)-\pazocal{P}_t(\xi,k)\, .
%\frac{1}{2}(\mathcal{W}_t^{-,+}(\xi,k)+\mathcal{W}_t^{+,-}(\xi,k))  \\\nonumber
% & = \frac{1}{2}(\mathcal{W}_t^{-,+}(\xi,k)+\mathcal{W}_t^{+,-}(\xi,k) - \mathcal{W}_t^{-,+}(\xi,-k)+ \mathcal{W}_t^{-,+}(\xi,-k))  \\\nonumber
%& = \frac{1}{2}(\mathcal{W}_t^{-,+}(\xi,k)+ \mathcal{W}_t^{-,+}(\xi,-k)) = \frac{1}{2}({W}_t(\xi,k)+ {W}_t(\xi,-k)) .
\end{align}
where we have employed the definitions in \eqref{eq:defHIPQ}.

For $\mathcal{U}^{11}$ and $\mathcal{U}^{12}$ the factor 
$1/\omega(k')$ complicates rewriting the result in terms of 
$\mathcal{W}$.   However, it is possible to go back to the scheme used for estimating \eqref{eq:flowterm}
and exploit the regularity of the smoothing function to find out the dominant contribution.
We begin by rewriting the sum over $k'$ as a convolution: 
\begin{align}
 & \int_{\Lambda_L^*} \rmd k' \, \frac{\rme^{2 \pi \ci x (k+k')}}{\omega(k')} \E[\widehat \psi_t(k',\sigma') \widehat \psi_t(k,\sigma)] 
 %\nonumber \\ & \quad
 = \sum_{x'\in \Lambda_L} 
\widetilde{\omega^{-1}}(x-x') \rme^{2 \pi \ci (x-x')\cdot k} Y_t^{\sigma',\sigma}(x',k)
% \widetilde{\omega^{-1}}(x-x') \int_{\Lambda_L^*} \rmd k' \, \rme^{2 \pi \ci x' (k+k')} \E[\widehat \psi_t(k',\sigma') \widehat \psi_t(k,\sigma)]
\end{align}
where $\widetilde{\omega^{-1}}(y) = \int_{\Lambda_L^*} \rmd k' \, \rme^{2 \pi \ci y \cdot k'} \omega(k')^{-1}$ 
is the inverse Fourier transform of $1/\omega$.  Therefore,
\begin{align}
 & \sum_{x \in \Lambda_L} \varphi(\xi-x) \int_{\Lambda_L^*} \rmd k' \, \frac{\rme^{2 \pi \ci x (k+k')}}{\omega(k')} \E[\widehat \psi_t(k',\sigma') \widehat \psi_t(k,\sigma)] 
 \nonumber \\ & \quad
 = \sum_{y,x' \in \Lambda_L} \varphi(\xi-x'-y)
\widetilde{\omega^{-1}}(y) \rme^{2 \pi \ci y\cdot k} Y_t^{\sigma',\sigma}(x',k)
 \nonumber \\ & \quad
 = \frac{1}{\omega(-k)}\rme^{-\ci t \omega(k)(\sigma'+\sigma)} \mathcal{W}_t^{\sigma',\sigma}(\xi,k) 
 \nonumber \\ & \qquad
 +
 \sum_{x',y \in \Lambda_L} \left(\varphi(\xi-x'-y)-\varphi(\xi-x')\right)
\widetilde{\omega^{-1}}(y) \rme^{2 \pi \ci y\cdot k} Y_t^{\sigma',\sigma}(x',k)
 \, .
\end{align}
Here we use that 
$\varphi(\xi-x'-y)-\varphi(\xi-x') = -y\cdot\nabla \varphi(\xi -x') + O((y/R)^2)$ and, with the small caveat about the difference between Fourier transforms on a finite and an infinite lattice explained in Section \ref{sec:derivationeqU_gammalarge} at \eqref{eq:discretederivcorrect}, we obtain
\begin{align}
 & \sum_{x \in \Lambda_L} \varphi(\xi-x) \int_{\Lambda_L^*} \rmd k' \, \frac{\rme^{2 \pi \ci x (k+k')}}{\omega(k')} \E[\widehat \psi_t(k',\sigma') \widehat \psi_t(k,\sigma)] 
 \nonumber \\ & \quad
 = \frac{1}{\omega(k)}\rme^{-\ci t \omega(k)(\sigma'+\sigma)} \mathcal{W}_t^{\sigma',\sigma}(\xi,k)
% \nonumber \\ & \qquad
 - \ci \frac{1}{\omega(k)^2} \frac{\nabla\omega(k)}{2 \pi} \cdot \nabla_{\!\xi} \mathcal{W}_t^{\sigma',\sigma}(\xi,k)
 \rme^{-\ci t \omega(k)(\sigma'+\sigma)} + O(R^{-2})
 \, .
\end{align}
Applied in the definitions of $\mathcal{U}^{11}$ and $\mathcal{U}^{12}$, we thus find
\begin{align}\nonumber
\mathcal{U}^{11}_t(\xi,k)& =\frac{1}{\omega(k)^2} (\pazocal{H}_t(\xi,k)+\pazocal{P}_t(\xi,k))
-\ci \frac{1}{\omega(k)^3} v(k)\cdot(\nabla_{\!\xi}\pazocal{H}_t(\xi,k)+\nabla_{\!\xi}\pazocal{P}_t(\xi,k)) + O(R^{-2})\, ,\\
\mathcal{U}^{12}_t(\xi,k)& = -\ci
\frac{1}{\omega(k)} (\pazocal{I}_t(\xi,k)-\pazocal{Q}_t(\xi,k))
-\frac{1}{\omega(k)^2} v(k)\cdot(\nabla_{\!\xi}\pazocal{I}_t(\xi,k)-\nabla_{\!\xi}\pazocal{Q}_t(\xi,k)) + O(R^{-2})\, .
\end{align}

In Section \ref{sec:stationary}, we found that for stationary homogeneous systems $0=\pazocal{I}=\pazocal{P}=\pazocal{Q}$
and $\pazocal{H}$ is a real constant.  We could now repeat the analysis by including the derivative terms, and obtain also the magnitude of $O(R^{-1})$ corrections for near stationary systems.  For instance, then 
\begin{align}\label{eq:Iinhomog}
\partial_t \pazocal I_t (\xi,k) + v(k)\cdot \nabla_{\!\xi} \pazocal{H}_t(\xi,k) = -\gamma \pazocal{I}_t(\xi,k)\, ,
\end{align}
thus near stationarity $$ \pazocal{I}_t(\xi,k) \simeq -\gamma^{-1} v(k)\cdot \nabla_{\!\xi} \pazocal{H}_t(\xi,k)= O(R^{-1}).$$
The analysis of the magnitude of the other terms is, however, more involved.   If one concentrates on the scaling and ignores possible regularity issues,
it is possible to reproduce the computations in Section \ref{sec:stationary}, and in general one should have then
$\pazocal{P},\pazocal{Q} = O(R^{-2})$ and $\bar{\mathcal{C}}[\pazocal{H}_t] = O(R^{-2})$, implying $\pazocal{H}_t(\xi,k)=E_t(\xi)+O(R^{-2})$.  (We have sketched some details of the derivation in Appendix \ref{sec:quasistat}.)
Whenever this is the case, the particle correlations satisfy
\begin{align}
\mathcal{U}^{11}_t(\xi,k)& =\frac{1}{\omega(k)^2} E_t(\xi) 
-\ci \frac{1}{\omega(k)^3} v(k)\cdot\nabla_{\!\xi}E_t(\xi) + O(R^{-2})\, ,\\
\mathcal{U}^{12}_t(\xi,k)& = \ci \gamma^{-1}
\frac{1}{\omega(k)} v(k)\cdot \nabla_{\!\xi} E_t(\xi) + O(R^{-2})\, , \\
\mathcal{U}^{21}_t(\xi,k)& = - \ci \gamma^{-1} \frac{1}{\omega(k)} v(k)\cdot \nabla_{\!\xi} E_t(\xi) + O(R^{-2})
\, ,
\\\nonumber
\mathcal{U}^{22}_t(\xi,k)& = E_t(\xi) + O(R^{-2})\, .
\end{align}
Thus, written in a matrix form,
\begin{align}\label{eq:U_semifinal}
\mathcal{U}_t(\xi,k)= 
 E_t(\xi)
\begin{pmatrix}
\omega(k)^{-2} &  0 \\
 0 & 1
\end{pmatrix} - \ci \frac{1}{\omega(k)} v(k)\cdot \nabla_{\!\xi} E_t(\xi) 
\begin{pmatrix}
\omega(k)^{-2} &  -\gamma^{-1}\\ 
 \gamma^{-1}  & 0
\end{pmatrix}  + O(R^{-2}).
\end{align}

Let us point out that this result could have been derived from the Boltzmann equation discussed in Section \ref{sec:conductivity} by assuming 
that $\pazocal{P}$, $\pazocal{Q}$ and $\im W_t(\xi,k)$
are of lower order, namely $O(R^{-2})$.  Since then $W_t(\xi,k)^*=W_t(\xi,k) + O(R^{-2})$, the definitions directly imply that
$\pazocal{H}_t$ is equal to the part of $W_t$ even in $k$, and
$\pazocal{I}_t$ is equal to the part odd in $k$.  Hence the result $W_t(\xi,k) =  E_t(\xi)-\gamma^{-1} v(k) \nabla_{\xi} E_t(\xi)+O(R^{-2})$
implies precisely that $\pazocal{H}_t(\xi,k)=E_t(\xi)+O(R^{-2})$ and 
$\pazocal{I}_t(\xi,k) = -\gamma^{-1} v(k)\cdot \nabla_{\!\xi} E_t(\xi) + O(R^{-2})$.  This, together with $\pazocal{P},\pazocal{Q} = O(R^{-2})$,
suffices to give the form \eqref{eq:U_semifinal}
for the covariance matrix.

\section{Discussion}\label{sec:conclusions}

The kinetic theory of the velocity flip model 
discussed in Section \ref{sec:Kinetic} shows that from the point of view of phonons, the diffusive scale
relaxation on the level of the second order correlations is entirely described by the simple phonon Boltzmann equation 
(\ref{eq:WpBE}) for the polarization component $W_t=\mathcal{W}_t^{-,+}$ while the self-polarization component
$\mathcal{W}_t^{-,-}$ may be set to zero at that stage of the evolution.  The dominant contribution in this picture is given by
a local equilibrium term and the first order corrections are directly related to energy currents.

Transformed from phonon modes back to $(q,p)$-fields, these two terms yield the expansion (\ref{eq:U_semifinal}) for the 
spatially averaged correlation matrix.  A comparison with the earlier result derived using the explicit estimates,
given in (\ref{eq:U_gammalarge}), shows that the two dominant terms are identical.  The 
lattice kinetic temperature profile $T_t(x)$, $x\in \Lambda$, in (\ref{eq:U_gammalarge}) merely
needs to be replaced by its spatially
averaged version $E_t(\xi)$, $\xi\in \R$, in (\ref{eq:U_semifinal}) (note that the temperature is equal to the
energy density in this model at thermal equilibrium).

Also for $(q,p)$-fields the dominant correlations are determined by the local thermal equilibrium correlations.  However,
the first order corrections acquire a term which is not related to the current observable, namely an additional correction 
to the $(q,q)$-correlations.  As seen from the computations in Section \ref{sec:kin_corr}, this correction arises from
the convolution which transforms the phonon modes back to particle variables.  It is also evident 
that this correction is zero whenever the state is translation invariant, so we may interpret it as a correction arising 
from changes in the phonon eigenbasis related to the inhomogeneous energy distribution.

Moreover, the 
relation between the phonon modes and the original Hamiltonian variables---between
$\mathcal{W}$ and $\mathcal{U}$ in Section \ref{sec:kin_corr}---does not depend on how the harmonic Hamiltonian 
evolution is perturbed.  However, the kinetic theory collision operator will greatly depend on the perturbation
and hence it is not obvious that the self-correlation terms can be neglected in all models
relevant to transport of phonons in crystalline structures.  For instance, 
it would be of interest to check more carefully what happens in models related to real three-dimensional crystals where the 
perturbations are small nonlinearities in the potential and there can be many different dispersion relations, as well as
multidimensional phonon mode eigenspaces.

To avoid complications arising from boundary effects, we have considered here only 
energy transport in periodic particle chains.  Another commonly used setup is to use fixed boundary conditions
and attach two thermostats to each end of the chain.  The thermostats drive the ends towards thermal equilibrium
with some predetermined temperatures, and such a system is then expected to reach a steady state with a 
temperature profile which can be solved from the Fourier's law using the boundary conditions 
given by the thermostats.  Then at the steady state in the bulk, i.e., sufficiently far away from the boundary,
the system will have a temperature gradient $O(1/L)$ where $L$ is the length of the chain.  As the effect of the thermostats
to the dynamics is expected to 
remain concentrated to the boundary, the bulk dynamics should be well approximated by the dynamics of the 
periodic chain.  Therefore, as a consequence of the above results, we can now make a precise conjecture 
about the structure of the above nonequilibrium steady state correlations: the dominant local correlations are
determined by the value of temperature at the site but they exhibit a correction whose leading term
is proportional to the temperature gradient and has the structure derived above in (\ref{eq:U_gammalarge}).

Here we have compared two different schemes to study thermal transport in the velocity flip model:
the explicit estimates relying on the renewal equation and the kinetic theory from 
the spatially averaged Wigner function.
The comparison highlights the strengths and weaknesses of both approaches. Renewal equation and the pointwise estimates
are more sensitive to the local lattice dynamics and can detect for instance degeneracies which are
washed out by the spatial averaging in the other approach.  For instance, a chain with only next-to-nearest particle potential and an even number of particles will decouple into two non-interacting chains which thermalize independently from each other and 
might, for instance, reach different temperatures at equilibrium.
This is one of the reasons for the somewhat complicated condition---which fails in the above degenerate case---for the dispersion relation in \cite{L14} where uniform microscopic control was the goal.  

However, it is probably fair to assume that 
the explicit computations in \cite{L14} do not easily carry over to other models and the uniform control will remain a 
hard goal for most phonon systems.  
Although the spatial averaging can wash out relevant details from the dynamics,
it is the key to the separation of scales between transport and collisions in the kinetic theory computations in Section \ref{sec:Kinetic}.  
We assume there $\gamma^{-1}\ll R\lesssim L$, but some additional assumptions will likely be needed if one wishes to complete the missing details and prove rigorously that the conjectured behaviour actually occurs for the velocity flip model.  However, as the best one could hope for from such a computation in the velocity flip model would be a reproduction of the existing diffusion proofs, the extra effort would likely pay off only in other, more complicated, models such as particle chains with anharmonic perturbations.
% This scale separation property is quite general and could become useful in 
% further studies of other phonon systems.

\appendix
\section{Computation of $ q(k)$ in \eqref{eq:tildeQk}
% $\tilde{Q}(k)$
} \label{app}

In this section we show the explicit computation which proves \eqref{eq:qkexplicit}. We look at 
\eqref{eq:qk}, i.e.
\begin{equation}\label{eq:qkbis}
 q(k)=2 \,\gamma\int_0^{\infty} \rmd t \, \widehat{A}_t(k)^{22} \partial_k \widehat{A}_t(k)^{21}.%=\frac{\partial_k \omega(k)}{ \gamma \omega(k)}\,.
\end{equation}
Here 
\begin{align*}\nonumber
& \widehat{A}_t(k)^{22}=\frac{\rme^{-\frac{\gamma \,t}{2}}}{\Omega}\left(-\frac{\gamma}{2}\sinh \Omega\,t +\Omega \cosh \Omega\, t \right),\\\nonumber
&  \displaystyle  \partial_k \widehat{A}_t(k)^{21}= \frac{\rme^{-\frac{\gamma \,t}{2}}\Omega'}{\Omega}\left(t\,\cosh \Omega\, t -\frac{\sinh \Omega \,t}{\Omega} \right).
\end{align*}
Therefore the integrand in \eqref{eq:qkbis}, i.e., $2 \,\gamma\, \widehat{A}_t(k)^{22} \partial_k \widehat{A}_t(k)^{21}$ reads
\begin{align*}\nonumber
%&2 \,\gamma\, \widehat{A}_t(k)^{22} \partial_k \widehat{A}_t(k)^{21}\\&
&2 \,\gamma \frac{\rme^{-\gamma t}\Omega'}{\Omega^2}\left(-\frac{\gamma \, t}{2}\sinh \Omega\, t\, \cosh \Omega\, t +\frac{\gamma}{2\,\Omega} \sinh ^2 \Omega \,t +\Omega\,t \cosh^2 \Omega \,t-\cosh \Omega \,t\,\sinh \Omega \,t\right)\\\nonumber
&=2 \,\gamma \frac{\Omega'}{\Omega^2}\left[-\frac{\gamma \, t}{8}\left(\rme^{-t(\gamma-2\,\Omega)}-\rme^{-t(\gamma+2\,\Omega)}\right)+\frac{\gamma}{8\,\Omega}\left(\rme^{-t(\gamma-2\,\Omega)}+\rme^{-t(\gamma+2\,\Omega)}-2\rme^{-t\,\gamma}\right)\right]\\\nonumber
&\quad +2 \,\gamma \frac{\Omega'}{\Omega^2}\left[\frac{\Omega \, t}{4}\left(\rme^{-t(\gamma-2\,\Omega)}+\rme^{-t(\gamma+2\,\Omega)}+2\rme^{-t\,\gamma}\right)-\frac{1}{4}\left(\rme^{-t(\gamma-2\,\Omega)}-\rme^{-t(\gamma+2\,\Omega)}\right)\right].
\end{align*}
where the ``prime'' denotes the derivative with respect to $ k$.
Once we integrate with respect to the time variable, we obtain
\begin{align*}
q(k)&=2 \,\gamma \frac{\Omega'}{\Omega^2}\left[-\frac{\gamma}{8}\left(\frac{1}{(\gamma-2\,\Omega)^2}-\frac{1}{(\gamma+2\,\Omega)^2}\right)+\frac{\gamma}{8\Omega}\left(\frac{1}{(\gamma-2\,\Omega)}+\frac{1}{(\gamma+2\,\Omega)}-\frac{2}{\gamma}\right)\right]\\&
\quad +2 \,\gamma \frac{\Omega'}{\Omega^2}\left[\frac{\Omega}{4}\left(\frac{1}{(\gamma-2\,\Omega)^2}+\frac{1}{(\gamma+2\,\Omega)^2}+\frac{2}{\gamma^2}\right)-\frac{1}{4}\left(\frac{1}{(\gamma-2\,\Omega)}-\frac{1}{(\gamma+2\,\Omega)}\right)\right]\\&
%=2 \,\gamma \frac{\Omega'}{\Omega}\left[-\frac{\gamma^2\Omega}{(\gamma^2-4\,\Omega^2)^2}%+\frac{\Omega}{(\gamma^2-4\,\Omega^2)}
%+\frac{\Omega}{4}\left(\frac{2(\gamma^2-4\,\Omega^2)^2+\gamma^2(\gamma-2\,\Omega)^2+\gamma^2(\gamma+2\,\Omega)^2}{\gamma^2(\gamma^2-4\,\Omega^2)^2}\right) %- \frac{\Omega}{(\gamma^2-4\,\Omega^2)}
%\right]\\&
%=2 \,\gamma \frac{\Omega'}{\Omega}\left[-\frac{\gamma^2\Omega}{(\gamma^2-4\Omega^2)^2}+
%\frac{\Omega}{4}\left(\frac{2(\gamma^2-4\Omega^2)^2+\gamma^2(\gamma-2\Omega)^2+\gamma^2(\gamma+2\Omega)^2}{\gamma^2(\gamma^2-4\Omega^2)^2}\right)\right]
=2 \,\gamma \frac{\Omega'}{\Omega^2}\left[-\frac{\gamma^2\Omega}{(\gamma^2-4\,\Omega^2)^2}+
\frac{\Omega}{\gamma^2}\left(\frac{\gamma^4-2\,\Omega^2\gamma^2+8\,\Omega^4}{(\gamma^2-4\,\Omega^2)^2}\right)\right]
%=2 \,\gamma \frac{\Omega'}{\Omega}\left[-\frac{\gamma^2\Omega}{(\gamma^2-4\,\Omega^2)^2}+
%\Omega \frac{\gamma^2-2\,\Omega^2}{(\gamma^2-4\,\Omega^2)^2}+\frac{8\,\Omega^5}{(\gamma^2-4\,\Omega^2)^2}\right]\\&
%=4  \,\Omega' \Omega \,\frac{4\,\Omega^2-\gamma^2}{\gamma\,(\gamma^2-4\,\Omega^2)^2} 
\\\nonumber
& =-\frac{4  \,\Omega' \Omega}{\gamma\,(\gamma^2-4\,\Omega^2)}.
\end{align*}
Since $\Omega = (\gamma /2) \sqrt{1 - (2 \omega(k)/\gamma)^2} $ it results that $\Omega' =-\frac{\omega\,\omega'}{\Omega}.$
By inserting the explicit expression for $\Omega'$ in the previous computation we get
$$
q(k)= \frac{\partial_k \omega (k)}{\gamma\,\omega(k)}.
$$

\section{Basic properties of lattice averaging kernels}\label{sec:latticeavk}

In Section \ref{sec:Wignernew}, we referred to ``lattice averaging kernels''
which were understood as convolution sums constructed using the kernel functions
\begin{align}%\label{eq:test_function}
 \varphi(\xi)= \frac{1}{R^d} \sum_{n \in \Z^d} \phi\!\left(\frac{\xi - L n}{R}\right)\, , \qquad \xi \in \R^d\, ,
\end{align}
for some given $L,R>0$.
These kernels are determined via the function $\phi :\R^d\to \R$ which we assume to satisfy all of the following 
conditions
\begin{jlist}
 \item $\phi$ is a Schwartz test function, i.e., $\phi \in \mathcal{S}(\R^d)$.
 \item $\FT{\phi}$ has a compact support.  Let $\rho_\phi>0$ be such that 
 $\FT{\phi}(p)=0$ whenever $|p|_\infty\ge \rho_\phi$.
 \item $\phi\ge 0$.
 \item $\int \!\rmd y \, \phi(y)=1$.
\end{jlist}
Since this construction could become useful in phonon models in higher dimensions, we write the results below for arbitrary $d\ge 1$,
keeping in mind that in the text they are applied with $d=1$.  The main difference comes from the fact that for $d>1$, the max-norm 
$|y|_\infty := \max_{1\le k\le d} |y_k|$ and the Euclidean norm $|y| := (y_1^2+y_2^2+\cdots + y_d^2)^{1/2}$ no longer give the same numbers.
We will mainly need the max-norm for the present lattice systems.

Let us show next that these assumptions guarantee the following properties for $\varphi$:
\begin{jlist}
 \item\label{it:pos} (positivity) $\varphi\ge 0$.
 \item\label{it:per} ($L$-periodicity) $\varphi(\xi+L m)=\varphi(\xi)$ for all $\xi\in \R^d$, $m\in \Z^d$.
 \item\label{it:normalc} (continuum normalization) $\int_{|\xi|_\infty \le L/2} \!\rmd \xi\, \varphi(\xi-\xi_0)=1$ for all $\xi_0\in \R^d$.
 \item\label{it:slowv} (slow variation) To every multi-index $\alpha$ there is a
 constant $C_\alpha$, which is independent of $R$ and $L$, such that 
 \begin{align}
  \left|\partial_\xi^\alpha \varphi(\xi)\right| \le R^{-|\alpha|} C_\alpha\, , \quad \text{ for all }\xi\in \R^d\, .
 \end{align}
 \item\label{it:normal} (lattice normalization) If $R\ge \rho_\phi$, we have 
 $\sum_{x\in \Lambda_L}\varphi(\xi+x)=1$ for all $\xi\in \R^d$.
 \item\label{it:fourier} (discrete Fourier transform) If $R\ge 2 \rho_\phi$, we have for all $k\in\Lambda_L^*$, $\xi\in \R^d$
 \begin{align}
 \sum_{x\in \Lambda_L} \varphi(\xi-x)  \rme^{-\ci 2\pi x\cdot k}
 =  \rme^{-\ci 2\pi \xi\cdot k} \FT{\phi}(-Rk)\, .
\end{align}
\end{jlist}
Hence, the constant $L$ determines the periodicity of the kernel and $R$ the scale of variation, in the sense
that each derivative of $\varphi$ will decrease the magnitude by $R^{-1}$.

The items \ref{it:pos} and \ref{it:per} are obvious consequences of the definition of $\varphi$ and the assumptions on $\phi$.
Item \ref{it:normalc} is derived by rewriting the sum over integrals as a single integral as follows:
\begin{align}
 \int_{|\xi|_\infty \le L} \!\rmd \xi\, \varphi(\xi-\xi_0) = 
 \frac{1}{R^d} \sum_{n \in \Z^d} 
  \int_{|\xi|_\infty \le L/2} \!\rmd \xi\, \phi\!\left(\frac{\xi + L n-\xi_0}{R}\right)
= \frac{1}{R^d} 
  \int_{\R^d} \!\rmd y\, \phi\!\left(\frac{y-\xi_0}{R}\right) = 1\, .
\end{align}

Item \ref{it:slowv} follows by taking the derivative inside the sum over $n$, and then noticing that the result
can be bounded by $R^{-|\alpha|}$ times a Riemann sum approximation of the integral $\int\! \rmd y\, |\partial^\alpha\phi(y)|$
which is finite since $\phi$ is a Schwartz function.

The lattice normalization condition and Fourier transform in items \ref{it:normal} and \ref{it:fourier} 
need slightly more effort.  Applying the definitions of $\varphi$ and of
the finite lattice $\Lambda_L$, we obtain for any $k\in\Lambda_L^*$, $\xi\in \R^d$:
\begin{align}
 \sum_{x\in \Lambda_L} \varphi(\xi-x)  \rme^{-\ci 2\pi x\cdot k}
 =  \sum_{x\in \Lambda_L} 
 \frac{1}{R^d} \sum_{n \in \Z^d} \phi\!\left(\frac{\xi-x - L n}{R}\right)
 \rme^{-\ci 2\pi (x+L n)\cdot k}
  =  
 \frac{1}{R^d} \sum_{m \in \Z^d} f(m)
\end{align}
where $f(y):=\phi\!\left(\frac{\xi-y}{R}\right)\rme^{-\ci 2\pi y\cdot k}$ is a Schwartz function.
The Fourier transform of $f$ is given by $\FT{f}(p)=R^d \rme^{-\ci 2\pi \xi\cdot (p+k)} \FT{\phi}(-R(p+k))$.
Therefore, by the Poisson summation formula,
\begin{align}
 \sum_{x\in \Lambda_L} \varphi(\xi-x)  \rme^{-\ci 2\pi x\cdot k}
 =   
 \frac{1}{R^d} \sum_{m \in \Z^d} \FT{f}(m)
 = \sum_{m \in \Z^d}\rme^{-\ci 2\pi \xi\cdot (m+k)} \FT{\phi}(-R(m+k))\, .
\end{align}
If $m\ne 0$, we have $|m+k|_\infty\ge |m|_\infty-|k|_\infty\ge \frac{1}{2}$, and thus $|-R(m+k)|_\infty\ge R/2$.
Hence, if $R\ge 2 \rho_\phi$, or $k=0$ and $R\ge \rho_\phi$,
all these points lie outside the support of $\FT{\phi}$, and thus only the ``$m=0$'' term may contribute to the sum.
This yields
\begin{align}\label{eq:discr_FT}
 \sum_{x\in \Lambda_L} \varphi(\xi-x)  \rme^{-\ci 2\pi x\cdot k}
 =  \rme^{-\ci 2\pi \xi\cdot k} \FT{\phi}(-Rk)\, .
\end{align}
In particular, if $k=0$, we have $\FT{\phi}(-Rk)= \FT{\phi}(0)=\int \!\rmd y \, \phi(y)=1$, and we obtain
$\sum_{x\in \Lambda_L} \varphi(\xi-x)=1$.  This completes the proof of both item \ref{it:normal} and item \ref{it:fourier}.

\section{Quasi-stationary inhomogeneous solutions}\label{sec:quasistat}

Here we want to show that $ \pazocal P_t = \pazocal Q_t = O(R^{-2})$, $ \pazocal I_t =O(R^{-1})$ and $ \pazocal H_t = E_t + O(R^{-2})$ as anticipated in Section \ref{sec:kin_corr}. Using the definitions \eqref{eq:defHIPQ} including the $ \xi$-dependence,
as well as the antisymmetry $v(-k)=-v(k)$, from \eqref{eq:fullBE} we deduce
\begin{align}\label{eq:H_appr}
\partial_t \pazocal H_t(\xi,k) &= -v(k)\nabla_{\xi} \pazocal I_t(k,\xi) + \bar{\mathcal{C}}[\pazocal H_t-\pazocal P_t](\xi,k) + O(R^{-2}) \\ \label{eq:I_appr}
\partial_t \pazocal I_t(\xi,k) &= -v(k)\nabla_{\xi} \pazocal H_t(k,\xi)-\gamma \pazocal I_t(\xi,k) + O(R^{-2}) \\ \label{eq:PQ_appr}
\partial_t \begin{pmatrix}
\pazocal P_t(\xi,k) \\
\pazocal Q_t(\xi,k)
\end{pmatrix} &= \pazocal L_{v}\begin{pmatrix}
\pazocal P_t(\xi,k) \\
\pazocal Q_t(\xi,k)
\end{pmatrix} + \begin{pmatrix}
\gamma \int_{\Lambda_L^* } \rmd q \pazocal P_t(q) - \bar{\mathcal{C}}[\pazocal H_t](\xi,k) \\
0
\end{pmatrix}
\end{align}
where
\begin{align*}
\pazocal L_{v} = \begin{pmatrix}
-\gamma & 2 \ci \omega(k)-v(k)\nabla_{\xi} \\
2 \ci \omega(k)-v(k)\nabla_{\xi} & -\gamma
\end{pmatrix}.
\end{align*}
Recall that $ \pazocal H_t, \pazocal I_t, \pazocal P_t$ and $\pazocal Q_t$ are $ L$-periodic in $ \xi$.  To solve \eqref{eq:H_appr}, \eqref{eq:I_appr} and \eqref{eq:PQ_appr} we look at the Fourier coefficients of those observables: 
\begin{align}\label{eq:H_hat}
\partial_t \widehat{\pazocal H}_t(n,k) &= -2 \ci \pi n L^{-1}v(k)  \widehat{ \pazocal I}_t(n,k) + \bar{\mathcal{C}}[\widehat{\pazocal H}_t-\widehat{\pazocal P}_t](n,k) + O(R^{-2}) \\
\partial_t \widehat{\pazocal I}_t(n,k) &= -2 \ci \pi n L^{-1}v(k) \widehat{ \pazocal H}_t(n, k)-\gamma \widehat{\pazocal I}_t(n,k) + O(R^{-2}) \label{eq:I_hat}\\
\partial_t
\begin{pmatrix}
\widehat{\pazocal P}_t(n,k) \\
\widehat{\pazocal Q}_t(n,k)
\end{pmatrix} &= \widehat{\pazocal L}_{v}\begin{pmatrix}
\widehat{\pazocal P}_t(n,k) \\
\widehat{\pazocal Q}_t(n,k)
\end{pmatrix} + \begin{pmatrix}
\gamma \int_{\Lambda_L^* } \rmd q \widehat{\pazocal P}_t(n,q) - \bar{\mathcal{C}}[\widehat{\pazocal H}_t](n,k) \\
0
\end{pmatrix}, \label{eq:PQ_hat}
\end{align}
where
\begin{align*}
\widehat{\pazocal L}_{v} = \begin{pmatrix}
-\gamma & 2 \ci( \omega(k)-\pi n L^{-1}v(k)) \\
2 \ci( \omega(k)-\pi nL^{-1} v(k)) & -\gamma
\end{pmatrix}
\end{align*}
and $ \widehat{\pazocal H}_t(n,k) = L^{-1}\int_{0}^L \rmd \xi \, \rme^{-2 \pi \ci L^{-1} n \cdot \xi} \pazocal H_t(\xi,k)$ with $ n \in \Z$ and analogously for $ \widehat{\pazocal I}_t,\widehat{\pazocal P}_t$ and $\widehat{\pazocal Q}_t$.

Assuming that the time derivative yields a contribution order $ O(R^{-2})$ we have
\begin{align}
\widehat{\pazocal I}_t(n,k) &= -\frac{2 \ci \pi n L^{-1} v(k)}{\gamma} \widehat{ \pazocal H}_t(n, k) + O(R^{-2}),
\end{align}
which implies that \eqref{eq:H_hat} becomes
\begin{align}\label{eq:H_hat2}
\bar{\mathcal{C}}[\widehat{\pazocal H}_t](n,k) &= \bar{\mathcal{C}}[\widehat{\pazocal P}_t](n,k) + \gamma^{-1} (2\pi n L^{-1} v(k))^2  \widehat{ \pazocal H}_t(n,k) + O(R^{-2}).
\end{align} 
Moreover, for \eqref{eq:PQ_hat} we have
\begin{align}\label{eq:PQ_hat2}
\begin{pmatrix}
\widehat{\pazocal P}_t(n,k) \\
\widehat{\pazocal Q}_t(n,k)
\end{pmatrix} &= -\widehat{\pazocal L}_{v}^{-1}
\begin{pmatrix}
\gamma \int_{\Lambda_L^* } \rmd q \widehat{\pazocal P}_t(n,q) - \bar{\mathcal{C}}[\widehat{\pazocal H}_t](n,k) \\
0
\end{pmatrix} + O(R^{-2}),
\end{align}
where
$$
\widehat{\pazocal L}_{v}^{-1} =- \frac{1}{\gamma^2 + 4(\omega(k)-\pi n L^{-1} v(k))^2} \begin{pmatrix}
\gamma & 2 \ci(\omega(k)-\pi n L^{-1} v(k)) \\
2 \ci (\omega(k)-\pi n L^{-1} v(k)) & \gamma
\end{pmatrix}.
$$
Combining \eqref{eq:H_hat2} and \eqref{eq:PQ_hat2} we get
\begin{align}\label{eq:P_final}
\widehat{\pazocal P}_t(n,k) = -\frac{( \pi n L^{-1} v(k))^2}{(\omega(k)-\pi nL^{-1} v(k))^2} 
\widehat{\pazocal H}_t(n,k)\, .
\end{align}
By the definition of the test function $ \varphi$ given in \eqref{eq:test_function}, $\widehat{\pazocal H}_t(n,k)$ is concentrated on values of $ n$ such that $ n/L$ is of order $O(R^{-1})$. 
In fact, from the definition of $ \pazocal H_t(\xi, k)$ we get the explicit form of $ \widehat{\pazocal H}_t(n,k)$:
\begin{align}\label{eq:H_expl}
\widehat{\pazocal H}_t(n,k) = \frac{1}{L} \int_0^L \rmd \xi \, \rme^{-2 \pi \ci L^{-1}n \cdot \xi} \sum_{x \in \Lambda_L} \varphi(\xi-x) V_t(x,k)
\end{align}
where
\begin{align*}
V_t(x,k)= \sum_{y \in \Lambda_L} \rme^{-2 \pi \ci y \cdot k} \E [\psi_t(x,-1) \psi_t(x+y, +1)+\psi_t(x,+1) \psi_t(x+y, -1)]\,.
\end{align*}
Thanks to \eqref{eq:discr_FT}, \eqref{eq:H_expl} becomes
\begin{align}\nonumber
\widehat{\pazocal H}_t(n,k) = & \frac{1}{L}  \int_{\Lambda_L^*} \rmd k' \, \widehat V_t(k',k) \widehat \phi(R k') \int_0^L \rmd \xi \, \rme^{-2 \pi \ci \xi \cdot (n L^{-1} -k')} \\
= &  \widehat V_t(n L^{-1},k) \widehat \phi(R n L^{-1})
\end{align}
where $ \widehat V_t(k',k):= \sum_{x \in \Lambda_L^*} \rme^{-2 \pi \ci x \cdot k'} V_t(x,k)$ and we used the fact that
$$
\frac{1}{L} \int_0^L \rmd \xi \, \rme^{-2 \pi \xi \cdot (n L^{-1} -k')} = \mathbb{1}(k' = nL^{-1})\, \ \ \mbox{for any } k' \in \Lambda_L^*.
$$
Therefore, $ nL^{-1} \in \Lambda_L^*$ and, since $ \widehat \phi$ has compact support (see assumption (2) in Appendix \ref{sec:latticeavk}), we get that $ \widehat \phi (RnL^{-1})=0$ whenever $ |nL^{-1}| \leq \rho_{\phi} R^{-1}$, from which the claim follows.

The fact that $ \widehat{ \pazocal H}_t(n,k)$ vanishes for $ |nL^{-1}| \geq O(R^{-1})$ indicates
that ${\pazocal P}_t(\xi,k)=O(R^{-2})$.  Then clearly ${\pazocal Q}_t(\xi,k)=O(R^{-2})$
and $\bar{\mathcal{C}}[{\pazocal H}_t](\xi,k)=O(R^{-2})$, thus implying also 
$\pazocal H_t = E_t + O(R^{-2})$.

\subsection*{Acknowledgements}

We thank Giada Basile, Mario Pulvirenti, and Herbert Spohn for useful discussions on the topic,
and the anonymous reviewers for their suggestions for improvements.
The work has been supported by the 
Academy of Finland
via the Centre of Excellence in Analysis and Dynamics Research (project 271983) 
and from an Academy Project (project 258302), and partially 
by the French Ministry of Education through the grant ANR (EDNHS).
We are also grateful to the Erwin Schr\"{o}dinger Institute (ESI), Vienna, Austria
for organization of a workshop and providing an opportunity for the related discussions.
Alessia Nota acknowledges support also through the \textit{CRC 1060 The mathematics
of emergent effects}
at the University of Bonn, that is funded through the German
Science Foundation (DFG).

% \newcommand{\utildir}[1]{../../../texstuff/#1}
% % .bst -file
% \bibliographystyle{\utildir{abunst_titles}}
% % .bib-files
% \bibliography{\utildir{myabbr},\utildir{mrabbrev},\utildir{allrefs}}
% \end{document}

\end{document}